	\documentclass[10pt]{article}

	\usepackage{amsmath}
	\usepackage{amssymb}

	\usepackage{graphicx,subfigure}

	\usepackage{cite}

	\usepackage{color}

	\usepackage[labelfont=bf,labelsep=period,justification=raggedright]{caption}

	\usepackage{lipsum}
	\makeatletter
	\renewcommand{\@biblabel}[1]{\quad#1.}
	\makeatother
	\date{}
	\newcommand{\eps}{{\varepsilon}}
	\newcommand{\D}[2]{\frac{d#1}{d#2}}
	\newcommand{\PD}[2]{\frac{\partial#1}{\partial#2}}
	\newcommand{\PDD}[3]{\frac{\partial^{#1}{#2}}{\partial{#3}^{#1}}}
\usepackage{color}

	\usepackage{multirow}
	
\begin{document}
	
\begin{flushleft}
	{\Large
	\textbf{Effect of Dedifferentiation on Time to Mutation Acquisition in Stem Cell-Driven Cancers}
	}

	
	Alexandra Jilkine$^{1,2}$,
	Ryan N. Gutenkunst$^{1,\ast}$
	\\
	\bf{1} Department of Molecular and Cellular Biology, University of Arizona, Tucson, AZ, USA
	\\
	\bf{2} Department of Applied and Computational Mathematics and Statistics,
	University of Notre Dame, Notre Dame, IN, USA
	\\
	$\ast$ rgutenk@email.arizona.edu
\end{flushleft}

	\section*{Abstract}
	Accumulating evidence suggests that many tumors have a hierarchical organization, with the bulk of the tumor composed of relatively differentiated short-lived progenitor cells that are maintained by a small population of undifferentiated long-lived cancer stem cells.
It is unclear, however, whether cancer stem cells originate from normal stem cells or from dedifferentiated progenitor cells.
	To address this, we mathematically modeled the effect of dedifferentiation on carcinogenesis.
	We considered a hybrid stochastic-deterministic model of mutation accumulation in both stem cells and progenitors, including dedifferentiation of progenitor cells to a stem cell-like state.
	We performed exact computer simulations of the emergence of tumor subpopulations with two mutations, and we derived semi-analytical estimates for the waiting time distribution to fixation.
Our results suggest that dedifferentiation may play an important role in carcinogenesis, depending on how stem cell homeostasis is maintained. 
	If the stem cell population size is held strictly constant (due to all divisions being asymmetric), we found that dedifferentiation acts like a positive selective force in the stem cell population and thus speeds carcinogenesis.
If the stem cell population size is allowed to vary stochastically with density-dependent reproduction rates (allowing both symmetric and asymmetric divisions), we found that dedifferentiation beyond a critical threshold leads to exponential growth of the stem cell population.
	Thus, dedifferentiation may play a crucial role, the common modeling assumption of constant stem cell population size may not be adequate, and further progress in understanding carcinogenesis demands a more detailed mechanistic understanding of stem cell homeostasis.

\section*{Author Summary}
Recent evidence suggests that, like many normal tissues, many cancers are maintained by a small population of immortal stem cells that divide indefinitely to produce many differentiated cells.
Cancer stem cells may come directly from mutation of normal stem cells, but this route demands high mutation rates, because there are few normal stem cells.
There are, however, many differentiated cells, and mutations can cause such cells to "dedifferentiate" into a stem-like state.
We used mathematical modeling to study the effects of dedifferentiation on the time to cancer onset.
We found that the effect of dedifferentiation depends critically on how stem cell numbers are controlled by the body. If homeostasis is very tight (due to all divisions being asymmetric), then dedifferentiation has little effect, but if homeostatic control is looser (allowing both symmetric and asymmetric divisions), then dedifferentiation can dramatically hasten cancer onset and lead to exponential growth of the cancer stem cell population. Our results suggest that dedifferentiation may be a very important factor in cancer and that more study of dedifferentiation and stem cell control is necessary to understand and prevent cancer onset.

\section*{Introduction}

	Most tissues consist of three classes of cells: stem cells, transit-amplifying progenitor cells, and differentiated cells. Multicellular organisms require a tight control of cell division to ensure a proper balance between these different cell populations. 
	The cancer stem cell (CSC) hypothesis states that tumors are also hierarchically organized, with a small sub-population of cancer  cells driving cancer growth \cite{reya_nature_review}.
Individual cell tracing studies of tumor development strongly support the cancer stem cell hypothesis in many (but not all) types of cancer \cite{Driessens_Nature2012,medema}, and identifying these cells in tissues is an ongoing goal in cancer research. Lineage studies find that malignant tumors contain more cancer stem cells compared to benign tumors and that cancers gradually lose their tissue-like hierarchical organization as they evolve from the benign to malignant \cite{Driessens_Nature2012}. 

	Cells escape proliferation control after acquiring a series of mutations in a multi-step process \cite{attolini-Michor}.
While some cancers may require only a few mutations \cite{Michor_Nowak_2006}, the number of required (driver) mutations in solid cancers is larger, with up to twenty driver mutations \cite{Beerenwinkel}. In order to accumulate this critical number of mutations during a lifetime, cells either have to be long-lived or the mutation rate has to be large \cite{Beckman_Loeb}. Stem cells have been proposed to be likely candidates for the initial cell of mutation due to their long lifetime and sustained self-renewal capacity \cite{reya_nature_review}. In addition to their long life span, stem cells are able to generate full lineages of differentiated cells, thereby perpetuating mutations through clonal expansion. Given known division and mutation rates, theoretical studies have argued that the necessary number of mutations for carcinogenesis cannot be obtained in the stem cell population on a reasonable time scale without assuming either significant selective advantage or elevated mutation rates \cite{attolini-Michor,Beckman_Loeb}.
	However, there is conflicting evidence as to how early in tumor development cancers acquire an elevated mutation rate \cite{Abdel-Rahman,Klein2006}
	and  several cancer genome sequencing studies have estimated mutation rates during cancer initiation to be normal for some  types of cancer \cite{Calabrese_2004, Jones_PNAS2008,wang_PNAS}.

	Although a stem cell may sustain the first oncogenic hit, subsequent alterations required for development of CSCs can occur in descendent progenitor cells \cite{Visvader}. Dysregulation of pathways involved in stem cell self-renewal may lead to progenitor cells acquiring a stem cell-like phenotype. It remains an open question whether cancer stem cells originate from stem cells that escape homeostasis or from dedifferentiated progenitor cells that acquire infinite proliferating potential \cite{Passegue}. There is significant evidence that dedifferentiation may play a role in establishment of some cancers. 
	In the hematopoietic system,
	it has been shown that leukemic stem cells can be generated from committed progenitor cells that acquire stem cell-like behavior \cite{Krivtsov}.
It has been suggested that acute myeloid leukemia (AML) is a progenitor disease, where a progenitor acquires abnormal self-renewal potential and``dedifferentiates'' to a stem cell-like state \cite{Goardon,Zhao}. Other myeloid leukemias such as CML (chronic myeloid leukemia) are thought of as stem-cell diseases \cite{Dingli_BMC_Biol}.
	However, although a hematopoietic stem cell is thought to be the cell of origin in the early phases of CML, in patients with CML blast crisis, granulocyte--macrophage progenitors are thought to acquire self-renewal capacity through a $\beta$-catenin mutation and emerge as the probable CSCs \cite{Jamieson}. 
Using mathematical modeling to investigate the likelihood of mutation occurring in a progenitor versus a stem cell is a continuing line of investigation \cite{Haeno_PNAS}. We treat the probability of a mutant progenitor cell acquiring stem cell-like state as a ``dedifferentiation'' rate, and we study how this parameter influences the time to carcinogenesis.
	We are primarily interested in whether dedifferentiation
	can speed up the time to tumor development
	in hierarchically organized cancers and in what rates of dedifferentiation are necessary for a noticeable effect.


	\subsection*{Prior Related Mathematical Modeling}

	Certain aspects of the cancer stem cell hypothesis have previously been addressed
	by mathematical models. It has been shown that having a hierarchical tissue design, where a small population of stem cells maintains a transient population of differentiating cells, may slow the accumulation of mutations and protect against cancer \cite{FrankIwasaNowak2003,Komarova-review, Wodarz_book}. 
	The question of whether genetic instability (resulting in hyperactive mutation rate) is an early or later event in mutation acquisition leading to cancer has been addressed by several groups (see \cite{attolini-Michor} for review).
	Most mathematical models find that the onset of genetic instability should be an early event, if at least some of the mutations are neutral. 
    However, sequencing suggests that the mutator  phenotype is expressed relatively late in cancer progression \cite{Klein2006}.

	Stem cell populations are typically small. Hence, the dynamics of mutant cells in the stem cell population are highly sensitive to stochastic fluctuations. A tumor begins with a single mutated cell, so there is a substantial chance of mutant extinction due to random events. Genetic drift and stochastic clonal extinction in stem cell lineages have been experimentally demonstrated for both normal tissue stem cells \cite{klein_development, lopez-garcia_science, snippert_cell}  and cancer stem cells \cite{Driessens_Nature2012}  in several tissue types. A deterministic model of mutation acquisition in stem cells will significantly underestimate the time to cancer establishment \cite{Dingli_Cell_Cycle}. Many models of mutation acquisition use a stochastic approach and are concerned with calculating time to emergence or fixation (or when the number of mutant cells reaches some threshold value used in diagnosis) of a mutant cell with fitness $r=1+s$ in a population of size $N_{sc}$. 

	The waiting time for cancer is often defined as the time until a particular number of  mutation events have occurred in at least one cell.
	Iwasa et al. \cite{Iwasa2004} considered a two stage Moran model and described conditions under which ``stochastic tunneling'' can occur. (In this phenomenon, cells with two mutations reach fixation before cells with one mutation reach fixation.) Durrett et al. \cite{Durrett2009} obtained asymptotic estimates of waiting times until a cell with $i$ mutations first appears under assumption of neutrality ($s =0$).
	These models typically consider a fixed population size \cite{Michor_Nowak_2006,Dingli_PLOS_CB, Dingli_PhysRevE,Dingli _BMC_Biol,Foo2011, Haeno_PNAS,Thalhauser,yatabe_PNAS}. The fixed population assumption is supposed to reflect homeostasis in the stem cell population, though how homeostasis is achieved is typically not addressed.
	Although the Moran model captures the stochastic nature of mutation acquisition, this type of model is not capable of describing mutations that lead to the change of stem cell division pattern that results in possible expansion of the stem cell pool, which in turn leads to tumor growth.
	Some recent models also consider mutation accumulation in exponentially growing cell populations \cite{Bozic, Durrett2011, Haeno_genetics, Tomasetti2012}.
	Beerenwinkel et al. \cite{Beerenwinkel} used the Wright-Fisher model with exponentially growing population size to look at effect of selection on the waiting time to cancer, and they predicted that the observed genetic diversity of colorectal cancer genomes can arise under a normal mutation rate (taken to be $u=10^{-7}$ per cell division) if the average selective advantage per mutation is on the order of 1\%. Similar calculations using a discrete branching process found $s=0.4$\% given $u=10^{-5}$ \cite{Bozic}. Note that increased mutation rates due to genetic instability would allow even smaller selective advantages during tumorigenesis, but neutral mutants ($s=0$) result in waiting times that are too long compared with disease incidence.
	Other groups have also concluded that for  normal mutation rates and neutral mutants, mutations in multiple genes in acquired hematopoietic disorders are most likely very rare events, as acquisition of multiple mutations typically requires  development times that are too long  compared to disease incidence \cite{Dingli_PhysRevE}. 

	Spencer et al.\cite{spencer_2006} and Ashkenazi et al.\cite{ashkenazi-neoplasia} have focused on the sequential order of mutations associated with increased rates of proliferation, decreased rate of death, increased mutation rate, and other hallmarks of cancer that must accumulate before emergence of cancer. The sequence of mutations with the shortest waiting time to getting all the necessary mutations is considered the most likely mutational pathway \cite{spencer_2006,Haeno_PNAS}.
	However, these models do not consider
	the possibility that dedifferentiation of  progenitor cells can affect the time to multiple mutation acquisition.

The dividing progenitor cell population has previously been described by multi-compartment ODE models, with cells moving between compartments as they age \cite{ashkenazi-neoplasia,marciniak-czochra,Johnston_PNAS}. Note that in these models the exact number of different stages of differentiation is ambiguous and does not exactly correspond to mitotic events, as cells may undergo more than one division in each compartment stage \cite{marciniak-czochra}.
	Most of these models of age-structured cell populations assume a stem cell proliferation rate 
	that is dependent on the total number of cells and thus incorporate negative feedback as a means of achieving homeostasis \cite{mahaffy_jtb,marciniak-czochra_siam}. These deterministic models have focused on mechanisms that could regulate cell numbers that are necessary for homeostasis and efficient repopulation.
	We use a similar mathematical approach to model the progenitor population as \cite{marciniak-czochra_siam}, but we couple it to stochastic dynamics in the stem cell compartment.

	Some studies have previously considered the impact of the asymmetry of cell division on stem cell dynamics \cite{Dingli_PLOS_CB,Tomasetti2010}. Suppose that upon division a stem cell can produce zero, one, or two stem cells with probabilities $\alpha_D$, $\alpha_A$, and $\alpha_S$, respectively (Fig.~\ref{Figure1:schematic}A). The mean number of stem cell offspring is given by $\alpha_A+2 \alpha_S$. 
	In that case, the stem cell population is best described by a branching process with the expected number of cells at time $t$ given by $(\alpha_A+2 \alpha_S)^t$.
	However, a branching process either goes extinct or undergoes exponential growth, and thus it cannot capture stem cell dynamics at equilibrium. One solution is to use a conditional branching process \cite{Ewens_book}, where the probabilities for a branching process are conditioned to the total population size remaining constant by an unspecified sampling mechanism (i.e., assuming that the stem cell population remains in homeostasis), and this approach has been applied to stem cell dynamics \cite{yatabe_PNAS}.
We consider both fixed and time-varying but bounded stem cell population size in our models.

	\subsection*{Our Modeling Approach}
	We use mathematical modeling to study how the possibility of ``dedifferentiation'' of mutant progenitor cells into a stem cell-like state affects the waiting time to carcinogenesis.
	Dividing progenitor cells have large growing populations, so we use a deterministic model to describe their evolutionary dynamics. 
	Discretizing the progenitor cell population based on the number of divisions a cell has completed, we obtain an age-structured partial-differential-equation model for the number of differentiated cells of age $a$ at time $t$. 
	This modeling approach is based on the assumption that maturation of progenitor cells is a continuous process which progresses with a constant velocity.
	For stem cell populations, stochastic effects are important, because the proliferating stem cell population is typically small. 
	We use a stochastic model for stem cell dynamics as a boundary condition to the PDE governing differentiated cell expansion (Fig.~\ref{Figure1:schematic}B and C.) There is also feedback from the deterministic progenitor population to the stochastic stem cell population as a rate of ``dedifferentiation''.

To assess the effect of dedifferentiation on time to carcinogenesis, we consider models for stem cell dynamics with both fixed and variable stem cell numbers (Fig.~\ref{Figure1:schematic}D).
	We assume neutral fitness of mutant stem cells, with the proliferation advantage of the mutant phenotype appearing only in the progenitor stage, in line with what is known for some cancers \cite{Dingli_2010, Dingli_PNAS}.
	The main questions we address are:
	\begin{enumerate}
	\item What is estimated time to carcinogenesis (acquisition of $M$ mutations) in stem cell-driven cancers if dedifferentiation from the progenitor population is allowed?
	\item What magnitude of dedifferentiation rate is needed to significantly shorten time to cancer acquisition? Will dedifferentiation still change the waiting time to cancer if homeostasis in the stem cell population is maintained (population size remains constant) or does homeostasis need to be lost?
\item What is the effect of symmetric division of stem cells, which leads to a non-constant stem cell population size? Do stochastic fluctuations in the size of the stem cell pool speed up time to malignancy compared to a constant stem cell population size? 
	\end{enumerate}
Our general compartment model can be applied to different tissues, such as colonic crypts, mammary cells, and hematopoiesis.

	\section*{Models}

	\subsection*{Progenitor Cells}

	We assume that when progenitor transit-amplifying cells carrying $i$ mutations  divide, they produce progentitor cells of the same maturity stage, obtaining the linear PDE
	\begin{equation}\label{equation:Gentry}
	\PD{p_i}{t}+\frac{da}{dt} \PD{p_i}{a}=(\sigma(a)-\mu(a))p_i.
\end{equation}
	where $p_i(a,t)$ is the progenitor cell density of age $a$ at time $t$, $\sigma(a)$ is the age-dependent proliferation rate, and $\mu(a)$ is the age-dependent mortality rate. 
    We assume that the rate of maturation $\frac{da}{dt}$ does not depend on age $a$ and, without loss of generality, set it equal to $1$. 
	Similar age-structured population equations have been previously studied, 
    with focus on the regulatory feedback mechanisms that are necessary for homeostasis and structural stability of the steady state solution \cite{Johnston_PNAS, marciniak-czochra_siam}.   
	
    Extending Eq.~\eqref{equation:Gentry} to account for mutations between multiple subpopulations of progenitor cells (Fig.~\ref{Figure1:schematic}C) we obtain
\begin{subequations}\label{equation:Gentry_mutation}
	\begin{align}
	\PD{p_0}{t}+\PD{p_0}{a}&=((1-u^{*})\sigma_0(a)-\mu_0(a))p_0,\\
	\PD{p_1}{t}+\PD{p_1}{a}&=((1-u^*)\sigma_1(a)-\mu_1(a))p_1+u^* \sigma_0(a)p_0, \cdots\\
	\PD{p_k}{t}+\PD{p_k}{a}&=(\sigma_k(a)-\mu_k(a))p_k+u^* \sigma_{k-1}(a)p_{k-1}.
	\end{align}
	\end{subequations}
	Here $u^*$ is the mutation rate per cell per unit time and $p_i(a,t)$ is the number of progenitor cells of ``age'' $a$ from the subpopulation with $i$ mutations. 
    We assume $0 \leq i \leq M$, and no back mutation is allowed. 

	We assume that there is a separate stem cell population that give rise to newly born differentiated cells $p_i(0,t)$ that serves as a boundary conditions to the PDE system in Eq.~\eqref{equation:Gentry_mutation}.
Let $n_i(t)$ be the number of stem cells with $i$ mutations at time $t$.
Let $\alpha_{D,i}$ be the probability of a symmetric division that gives rise to two differentiated cells, $\alpha_{A,i}$ be the probability of an asymmetric division that gives rise to one stem cell and one differentiated cell, and $\alpha_{S,i}$ be the probability of a symmetric division that gives rise to two stem cells.
Then
\begin{equation}\label{BC1}
p_i(0,t)=(2\alpha_{D,i}+\alpha_{A,i})n_i(t).
\end{equation}

If we neglect mutation, the steady state solutions of Eq.~\eqref{equation:Gentry_mutation} have the form 
\begin{equation}
p_i(a,t)= \alpha n_i(t-a) e^{r_i(a)},
\end{equation}
where $\alpha=2\alpha_{D,i}+\alpha_{A,i}$ is the average number of stem cells of type $i$ produced per division and $r_i(a)=\int_{0}^{a} (\sigma_i(s)-\mu_i(s))\, ds$ is the age-dependent growth rate of the differentiated cell population (Supplemental Text). 
Hence, the long-term age distribution is largely determined by the functional forms of the differentiated cell birth and death rates {(Supplemental Fig.~S1 and S2.) 
Altered birth and death rates due to mutations can result in mutant subpopulations growing to higher plateaus in size, but the final population size will be bounded.
Our PDE system can be easily modified to have a maximal carrying capacity $K_i$ for each sub-population.
This does not qualitatively change the age distribution of progenitor cells (Supplemental Fig.~S1) and does not significantly affect the fraction of $i$-mutation cells in the total progenitor population (Supplemental Fig.~S3), so we do not consider it further.

	To mimic a maturity switch for cellular proliferation and death, we took the proliferation and death rates of differentiated cells per unit time to be
\begin{subequations}\label{equations:birth_death}
\begin{align}
\sigma(a)&=\frac{b}{2}\left(1 - \tanh (\rho_b(a-\omega_b))\right)\text{, and}\\
\mu(a) &= \frac{d}{2}(1 + \tanh (\rho_d(a-\omega_d))).
\end{align}
\end{subequations}
	Here $b$ and $d$ are the maximal proliferation and death/removal rates of progenitor cells.
	The age at which the proliferation switch occurs (i.e., half the progenitor cells stop dividing) is given by $\omega_b$, and the steepness of the proliferation switch is determined by $\rho_b$. 
    Similarly, the age at which half the cells begin to undergo apoptosis is given by $\omega_b$, and the steepness of the death switch is controlled by $\rho_d$. 
    If $\omega_b<\omega_d$, then differentiated cells between the ages of $\omega_b$ and $\omega_d$ are not replicating (senescent).  Note that setting either of these values to zero  results in a uniform rate of birth/death.
    Effects of varying proliferation/death parameters are shown in Supplemental Fig.~S2. The parameters governing proliferation, in particular $b$ and $\omega_b$, have much larger influence on the final differentiated cell population size than parameters governing death/removal. The steepness of the switch does not change the age distribution..
   
   \subsection*{Stem Cells}
	\subsubsection*{Constant Stem Cell Population Size}
	To model the evolutionary dynamics of a stem cell population under strict homeostasis (resulting in a fixed stem cell population size), we used the Moran stochastic process for $M+1$ different types, with mutations between types and neutral fitness \cite{Ewens_book}.
	Let the number of individuals carrying each possible number of mutations be given by {$\mathbf{n}=(n_0,n_1,...,n_M)$, where $\sum_{i=0}^M n_i=N_{sc}$.
	We considered two versions of this model, with and without dedifferentiation.
In both cases, we assumed that each stem cell divides, on average, every $T_{gen}$ chronological time units.
Thus, in a population of size $N_{sc}$, the average time between divisions was  $T_{gen}/N_{sc}$.

	In the first model, no dedifferentiation of progenitor cells was possible.
	Every $T_{gen}/N_{sc}$ time units, a single randomly chosen stem cell $j$ was removed and one cell $i$ was born with probability given by 
	\begin{equation}\label{Moran_generalized}
	P(\mathbf{n} \to \mathbf{n}+ \mathbf{e_i} -\mathbf{e_j})=\frac{n_j}{N_{sc}} \left[
	\sum_{h=0}^{M}m_{i,h} \frac{n_h}{N_{sc}} \right],
	\end{equation}
	where $m_{i,h}$ is the probability of changing to type $i$ from type $h$ per replication event, and $\mathbf{e_j}$ is a unit vector with 1 in the $j$th column.
	We considered a linear cascade of mutations in which the mutation matrix 
    $[m_{i,h}]$ 
    is given by
	\begin{equation}\label{mutation_matrix_cascade}
    \begin{bmatrix}
	1-u_1 & 0 & 0 & \cdots & 0\\
	u_1& 1-u_2 & 0 & \cdots & 0\\
	0& u_2 & 0 & \cdots & 0\\
	\vdots & \vdots & \vdots & \ddots & \vdots\\
	0 & 0 & 0 & \cdots & 1\\
	\end{bmatrix}.
	\end{equation}

We also considered a version of the model in which dedifferentiation of two-mutation differentiated cells was allowed, but the total stem cell population size remained fixed.
In this model, the probability of death of a $j$-mutation stem cell and birth of an $i$-mutation cell was given by
\begin{equation}\label{equation:dedifferentiation_fixedN}
P(\mathbf{n} \to \mathbf{n}+ \mathbf{e_i} -\mathbf{e_j})=\left(1-\varepsilon f \right) \frac{n_j}{N_{sc}} \left[\sum_{h=0}^{M}m_{i,h} \frac{n_h}{N_{sc}}\right]+  \delta_{i,2} \varepsilon \frac{\int_a p_2(a )\, da}{\sum_i \int_a p_i(a )\, da},
\end{equation}
where $\varepsilon$ is the proportion of cells in the stem cell pool that come from dedifferentiated cells at each replication event, and $\delta_{i,2}$ is the Kronecker delta function signifying that that only two-mutation progenitor cells can dedifferentiate.
Here
\begin{equation*}f=\frac{\int_a p_2(a )\, da}{\sum_i \int_a p_i(a )\, da}
\end{equation*}
is the proportion of two-mutation cells of all ages in the progenitor population, given that $p_i(a)$ is the density of differentiated cells of age $a$ carrying $i$ mutations.
We also considered a version of the model in which all progenitor cells, regardless of the number of mutations, could dedifferentiate (Supplemental Material).

	Because the Moran model has been studied extensively, we were able to use several existing results on the time to emergence and fixation of mutants.
	Let $\tau_M$ be the first time at which an individual carrying $M$ mutations emerges who will go on to fix in the population.
    We focus on the case $M=2$ because sequencing of acute myeloid leukemia genomes suggests that 
there are 2 driver mutations present~\cite{Ley_nature}. (See discussion for more details.) 
	Using branching process approximations, Durrett et al.~\cite{Durrett2009} calculated the waiting time for the Moran model under neutral drift of prior mutants.
For $M=2$, the probability density function for $\tau_2$ is given by
	\begin{equation}\label{eq:Durrett_modified}
	\phi(N_{sc} u_1 \sqrt{u_2 \rho_{fix}} \tau_2= t) \approx \frac{1-e^{-2t/\lambda}}{1+e^{-2t/\lambda}} \exp \left(-\int_0^t \frac{1-e^{-2\xi/\lambda}}{1+e^{-2\xi/\lambda}}\, d\xi \right),
	\end{equation}
	where $\lambda=N_{sc} u_1$, and $\rho_{fix}$ is the probability that a single mutant individual will fix in a population of size $N_{sc}$.
For neutral drift, $\rho_{fix}=1/N_{sc}$, and for weak selection
	\begin{equation}
	\rho_{fix}=\frac{1-\exp(-2s)}{1-\exp(-2s N_{sc})},
	\end{equation}
	where $M$-mutation cells have advantage $s \ll 1$ \cite{Ewens_book}. 

	The time to fixation of the subpopulation with $M=2$ mutations is a sum of two random variables: the time $\tau_2$ until appearance of a successful two-mutation cell (Eq.~\eqref{eq:Durrett_modified}) and the waiting time $\tau_{fix}$ from the time that mutant first appears until that mutant fixes \cite{Crow_book}. Note that this time is given in units of stem cell generation times $T_{gen}$. 
	The pdf of the total fixation time, $T_{fix}$, is given by the convolution
	\begin{equation}\label{eq:ultimate_prob_fixation}
	\phi( T_{fix})=\int_0^{T_{fix}} \phi(t) \tilde{\phi}(T_{fix}-t )\, d t
	\end{equation}
	
	of the probability density functions $\phi$ for time to first appearance of successful mutant and $\tilde{\phi}$ for time it takes that mutant to fix.
	$\tilde{\phi}$ can be obtained from the backward Kolmogorov equation for the probability of fixation $f$ of a gene with initial frequency $p_0$ before time $t$:

	\begin{equation}
	\PD{f(p_0,t)}{t}=\frac{p_0 (1-p_0)}{2 N_{sc}} \PDD{2}{f(p_0,t)}{p_0} + s p_0 (1-p_0) \PD{f(p_0,t)}{p_0},
	\end{equation}
	subject to boundary conditions $f(1,t)=1$ and $f(0,t)=0$ and initial condition $f(p_0,0)=\delta(p_0)$.
	Dividing $f$ by the ultimate probability of fixation $\rho_{fix}$ and differentiating with respect to $t$, we obtain the probability density function for $\tilde{\phi}$ as a function of initial allele frequency $p_0$ \cite{KimuraOhta1969}. 

\subsubsection*{ Variable Stem Cell Population Population Size}
	Our previous stem cell models couple birth and death events to keep the population size fixed, but we next decoupled these events to allow for a stochastically varying population size.
	For clarity, we refer to the total stem cell population size in this model by $S(t)$.
	Again assuming that the average replication time of a stem cell is $T_{gen}$, the interval between birth/death events in this stochastic stem cell model corresponds to $T_{gen}/S(t)$ time units in the progenitor cell model.
	We assume that homeostasis in the stem cell pool is maintained by control of cell fate upon division, and that each stem cell can produce zero, one, or two stem cell offspring. 
    For example, the possible offspring from a zero-mutation stem cell are: two differentiated zero-mutation cells with probability $\alpha_D$, one zero-mutation differentiated cell and one zero-mutation stem cell with probability $\alpha_A(1-u)$, one zero-mutation differentiated cell and one one-mutation stem cell with probability $\alpha_A u$, two zero-mutation stem cells with probability $\alpha_S (1-2u)$, and a one zero-mutation stem cell and one one-mutation stem cell with probability $2 u \alpha_{S}$.
(For simplicity we assume the probability of both offspring carrying new mutations to be negligible.)
In general, a stem cell carrying $i$ mutations can produce $k$ stem cell offspring carrying $j$ mutations with probability $P_i^j(k)$ given by
	\begin{equation}\label{multi-type GW unidirectional mutation}
P_i^i(0)=\alpha_D,\, P_i^i(1)=\alpha_A(1-u)+\alpha_S u,\, P_i^{i+1}(1)=\alpha_A u + \alpha_S u,\, P_i^i(2)=\alpha_S(1-2u).
	\end{equation}
With constant division probabilities $\alpha_D, \alpha_S$ and $\alpha_A$, this model is a rescaled Galton-Watson branching process, and the stem cell population either goes extinct in finite time or undergoes exponential growth when the mean number of stem cell progeny per cell, $m=2\alpha_S+ \alpha_A$, is greater than one \cite{Allen_book}.
To describe a stem cell population under homeostasis, the probabilities $\alpha_D$, $\alpha_A$, and $\alpha_S$ must depend on the total stem cell population size $S(t)$. 
We model a carrying capacity $K_i$ for stem cells carrying $i$ mutations, such that
\begin{equation}\label{density-dependent GW 2}
\alpha_{A,i}=\eta, \, \alpha_{S,i}=(1-\eta) \frac{K_i^\xi}{K_i^\xi+S(t)^\xi},\, \, \alpha_{D,i}=(1-\eta)\frac{S(t)^\xi}{K_i^\xi+S(t)^\xi}.
\end{equation}
As $\eta$ approaches one, $\alpha_A$ approaches one, so that most of the divisions that occur do not change the stem cell population size, and we recover the Moran process for all carrying capacity values.
The parameter $\xi$ controls strength of fluctuations about the carrying capacities; as $\xi$ increases, the fluctuations become smaller.
In this model, newly emerging mutants can still go extinct due to stochasticity, but the total population reaches a quasi-stationary regime at population size $K_i$.
Note that, because this is a quasi-stationary regime, eventually the stem cell population will go extinct, but the expected time to extinction is exponentially proportional to $K_i$ for $\xi=1$ and $\eta=0$ \cite{Parsons_genetics}. 
We chose the carrying capacity to be large enough that extinction of the stem cell population does not occur on a physiological timescale, and we initialized the stem cell population to be at carrying capacity with zero-mutation cells.

We considered two versions of the variable stem cell population size model. 
In the first case, no dedifferentiation was possible. 
In the second case, differentiated cells with $K$ mutations were allowed to dedifferentiate and re-enter the stem cell population. 
Let $\delta$ be the dedifferentiation rate per two-mutation progenitor cell per unit time.
Then the mean number of cells dedifferentiating in the interval $T_{gen}/S(t)$ between two stem cell replications is 
\begin{equation}
\lambda = \delta \frac{T_{gen} }{S(t)} P(t) = \delta \frac{T_{gen} }{S(t)} \int p_2(a,t )\, da.
\end{equation}
To introduce dedifferentiated cells into the stem cell population, at each replication event we calculate the mean number of dedifferentiated progenitor cells $\lambda$ and update the stem cell population:
\begin{equation}
[n_0, n_1, n_2] \to [n_0, n_1, n_2+\texttt{Poisson}(\lambda)],
\end{equation}
After the total stem cell number is updated, the probabilities of reproduction are re-calculated using Eq.~\eqref{multi-type GW unidirectional mutation}, and reproduction is carried out.

	\subsection*{Model Parameters}
	
	Parameters used are summarized in Table~\ref{parameter_summary}. We used parameter estimates from the human hematopoietic system because parameters for other cancers are less well known. We used $M=2$ as the number of necessary mutations to develop a cancerous phenotype.
	Although it has been estimated that for the human hematopoietic system there are 11,000-22,000 stem cells~\cite{abkowitz_blood}, which give rise to all blood and immune system cells, most of these cells are quiescent and only divide when body sustains an injury and needs to repopulate the hematopoietic system. Our model only considers actively dividing stem cells, which have been estimated by various methods to number around 100~\cite{Dingli_Cell_Cycle, Nash}.
	The entire actively dividing stem cell population has previously been modeled as  turning over once per year \cite{Dingli_Cell_Cycle}, but most recent estimates have an individual stem cell dividing every 25-50 weeks \cite{Catlin}. However, this is likely an over-estimate, as it is difficult to distinguish between actively dividing and quiescent stem cell populations. We assume that an  active stem cell divides every 20 weeks, which when multiplied by $N_{sc}$ results in active stem cell population turnover time of $T_{gen}=5$ weeks. (The entire stem cell population including quiescent cells turns over on a much longer timescale.)

	Whereas the size of the active hematopoietic stem cell pool is small, the number of progenitor cells such as granulocyte, erythroid, monocyte, and megakaryocyte colony-forming units (CFU--GEMM) 
    and granulocyte and monocyte colony-forming units (CFU--GM) 
    is much larger. There are approximately $10^5$ CFU--GEMM cells and $10^8$ CFU--GM cells \cite{Traulsen_bioessays}.
There are estimates that each CFU--GEMM may contribute to hematopoiesis for an average of 60 days (range of 40--340 days) and that it replicates at an average rate of once every 50 days (range of 35--285 days) \cite{Traulsen_bioessays}. We track the progenitor populations for $L=20$ weeks, and assume that their proliferative potential rapidly drops off after 10 weeks. The maximal proliferation and death rates, $b_i$ and $d_i$ were chosen so that 100 stem cells results in $10^{5}-10^{6}$  progenitor cells of all ages.

	Not much is known about the selective advantage $s$ provided by driver mutations for different cancer types, except that it is small ($r=1+s \approx 1$). 	Unless stated otherwise, we assume neutral fitness in the stem cell pool ($s=0$) in our stochastic models throughout the paper, to focus on the effect of dedifferentiation. We use a range of $s=0 \cdots 0.4$ for the progenitor cells in the deterministic model.
    
    Mutation estimates per cell division per gene range from about $10^{-7}$  in normal cells to $10^{-2}$ in case of chromosomal instability \cite{Nowak_book}. (Note that the rate of epigenetic change has been estimated to be orders of magnitude higher than that of genetic change and could also play a role in cancer initiation \cite{Calabrese_2004}.)
 A common value used in many mathematical models is a driver mutation rate of $u=10^{-5}$ per division, obtained by assuming a somatic mutation rate of $10^{-7}$ per gene, and about 100 genes that could be mutated to give same phenotype \cite{ashkenazi-neoplasia,Bozic}.
   In normal hematopoietic cells mutation rate has been measured as $u=10^{-6}$ per division \cite{araten}.

	Note that in the stochastic model, which considers every cell division, the mutation rate $u$ can be used as is, but using chronological time (i.e., weeks or months) means that this value should be multiplied by the average number of divisions per unit time to obtain $u^*$.
(Mutations that speed up the cell cycle will then speed up the apparent mutation rate per unit of chronological time in our progenitor model.)
	The expected number of doublings from $n_i$ stem to $p_i$ progenitor cells is $\log_2\left(p_i/n_i\right)+2 $ \cite{marciniak-czochra},
	 and the total number of progenitors cells of type $i$ is $p_i \approx \alpha n_i e^{\int r(a),\, da}$. Using values from Table~\ref{parameter_summary}, this results in $8-10$ cell divisions that take place over 10 weeks, so $u^* \approx u$ in equations~\eqref{equation:Gentry_mutation}.

\section*{Results}
The coupled system of stem cells and progenitor cells undergoing mutation and dedifferentiation we modeled is complex.
To disentangle the effects of different phenomena, we systematically built up the model.
We first considered the progenitor population alone.
We then considered the stem cell population alone, in models with strict and variable stem cell homeostasis.
Finally, we coupled the stem and progenitor populations through dedifferentiation.

\subsection*{Progenitor Population Alone}
We first considered whether mutation and reproduction in the progenitor population could by itself generate a sustained population of two-mutation cancerous cells.
We thus modeled a scenario in which no stem cell mutations occur, so the boundary condition to the progenitor population system in equations \eqref{equation:Gentry_mutation} is simply $(N_{sc},0, \cdots, 0)$.
Because selection in the progenitor population might favor mutants, we also assumed that progenitor cells with $i$ mutations have a proliferation rate $b_i = (1+s)^i b_0$ (Eq.~\ref{equations:birth_death}).
This yields a steady-state age distribution of normal and mutant progenitor cells (Supplemental
Fig.~2).

Fig.~\ref{Figure:competition plots} summarizes results for typical parameter values, showing that for $M=2$ mutant cells to be an appreciable fraction of the population, the mutation rate $u$ and proliferative advantage $s$ must both be unreasonably high.
This is true both if the total progenitor population can grow without bound (Fig.~\ref{Figure:competition plots}a) and if its growth is restricted (Fig.~\ref{Figure:competition plots}b). Similar findings are obtained if competition between progenitor subpopulations  is included in the model  (Supplemental
Fig.~3).
Consistent with previous work \cite{Dingli_PhysRevE,FrankIwasaNowak2003,Komarova-review},
these results show stem cell dynamics cannot be ignored in considering time to carcinogenesis, so we next considered stochastic models of the stem cell population.

\subsection*{Stem Cell Population Alone}
We next considered models of the stem cell population.
We began without dedifferentiation, so the dynamics are entirely governed by the stem cells.
In modeling cancer, the time to carcinogenesis can be defined as the time for a single $M$-mutation cell to emerge, the time for $M$-mutation cells to pass some threshold number or fraction, or the time for $M$-mutation cells to fix in the population.
If the mutation rate is low (such that $N_{sc} u \ll 1$), then all three definitions are similar, because the time to emergence of a successful $M$-mutation cell is long compared to the time from emergence to fixation.
However, there is large uncertainty regarding effective mutation rates in carcinogenesis (Table~1), so the assumption of low mutation rate may not always be valid, and we thus calculated times to fixation.

We began our stem cell modeling by considering fixed population size, corresponding to strict homeostasis.
In this constant $N_{sc}$ case, we could leverage several analytic results, with which our simulations agreed well.
Fig.~\ref{Figure:Moran_waiting_time}A shows a typical simulation. The full probability density distribution of time to fixation is given by Eq.~\eqref{eq:ultimate_prob_fixation} and agrees well with our simulations for high mutation rates (Fig.~\ref{Figure:Moran_waiting_time}B).
The time to emergence of a successful mutant is of order 1/($\sqrt{N_{sc}}u^{3/2}$) stem-cell generations (Eq.~\eqref{eq:Durrett_modified}).
For  normal mutation rates of $u=10^{-6}-10^{-8}$ per cell division, the mean time until emergence of a two-mutation cell is $10^8-10^{11}$ stem cell generations, which is very long even with a short stem cell generation time.

Because homeostasis is likely imperfect, we also considered a stochastically fluctuating stem cell population size.
We found that, without dedifferentiation, the distributions of times until fixation are very similar for models with and without fluctuations in the stem cell population size, as long as we condition on non-extinction of the stem cell population (Fig.~\ref{Figure:Moran_waiting_time}B).
This is true for a wide range of probabilities of asymmetric division $\eta$ and strengths of mean reversion $\xi$ (Eq.~\ref{density-dependent GW 2}).
This agrees with previous findings that demographic stochasticity does not alter fixation times of neutral mutants in a large population~\cite{Parsons_2008}, provided that the carrying capacities of the mutants are the same. 

Our results suggest that dynamics within either the progenitor or stem cell compartments considered separately do not result in carcinogenesis in the hematopoietic system on a realistic time-scale, provided that cancer-causing mutations occur at normal mutation rates, selection advantages relative to wild-type stem cells do not appear until $M=2$ mutations, and the stem cell population size is constant or varies stochastically around a carrying capacity.
We thus turned our attention to coupled model systems in which progenitor cells can dedifferentiate into stem cells.

\subsection*{Dedifferentiation With Constant Stem Cell Population Size}
For the coupled system, we first considered stem cell homeostasis caused by strict asymmetric division in the stem cell population, so the stem cell population size remains fixed.
To model dedifferentiation in this case, we built off the Moran model and assumed that when a stem cell dies and another enters the population, the new entrant comes from the two-mutation progenitor population with probability equal to $\varepsilon$ times the proportion of two-mutation cells in the progenitor population.
Otherwise the new stem cell comes from replication of another stem cell.
Roughly speaking, in this model the death of a stem cell leaves a opening in the niche, which can potentially be filled by a dedifferentiated progenitor cell.
The number of progenitor cells which can successfully dedifferentiate is controlled by the number of niche openings (stem cell deaths), not by the absolute number of progenitor cells.

Typical simulation results are shown in Fig.~\ref{Figure:dedifferentiation_stats_Moran_generations}A.
We found that dedifferentiation dramatically shortens the time to fixation of two-mutation cells (Fig.~\ref{Figure:dedifferentiation_stats_Moran_generations}B).
For small dedifferentiation rates $\varepsilon \lesssim 0.05$, we also saw good agreement between our simulations and a semi-analytical approximation for the time to fixation of two mutation cells with selective advantage $\varepsilon$ (Eq.~\eqref{eq:ultimate_prob_fixation}).
This agreement suggests that under strict stem cell homeostasis, dedifferentiation is effectively equivalent to a growth advantage for mutant stem cells.

Distributions of times to fixation of two-mutation stem cells are plotted as a function of both dedifferentiation rate $\varepsilon$ and mutation rate $u$ in Fig.~\ref{Figure:dedifferentiation_stats_Moran_generations}C.
Dedifferentiation had two major effects in this model: increasing the probability that an emergent two-mutation stem cell will fix and reducing the time between emergence and fixation.
Both of these effects act only after a two-mutation cell has been generated in the stem cell population.
(Recall that, as shown in Fig.~\ref{Figure:competition  plots}, the mutation rate and selective advantage must be unrealistically high for a nontrivial fraction of two-mutation progenitor cells to exist in the absence of underlying two-mutation stem cells.)
For all mutation rates $u$, the distribution of times to fixation was roughly constant for dedifferentiation rates $\eps \lesssim 1/N_{sc}$, consistent with population genetics theory that selection is only effective when the selection coefficient is greater than the reciprocal of the effective population size.
For small mutation rates $u$, increasing $\eps$ beyond this threshold only marginally shortened the total time to fixation.
This is because in this case the total time to fixation is dominated by the time for a successful two-mutation cell to emerge, and dedifferentiation only reduces this time by a factor of $1/\sqrt{\rho_{fix}}$ (Eq.~\ref{eq:Durrett_modified}), where $\rho_{fix}$ is the probability of a emergent two-mutation stem cell fixing.
Under neutrality $\rho_{fix} = 1/N_{sc}$, so for our model with $N_{sc} = 100$, dedifferentiation can shorten the time to emergence by at most a factor of 10. The dedifferentiation rate needed to significantly change this waiting time scales linearly with $N_{sc}$ (Supplemental Fig.~4C}). Hence, for larger stem cell population sizes, a small dedifferentiation rate would have a larger effect.
For high mutation rates $u$, the effect of dedifferentiation is more dramatic, because the time from emergence to fixation of two-mutation cells, which dedifferentiation also shortens, is comparable to the time to emergence (Fig.~\ref{Figure:dedifferentiation_stats_Moran_generations}D).

The model considered in Fig.~\ref{Figure:dedifferentiation_stats_Moran_generations} assumes that only two-mutation progenitor cells can dedifferentiate.
We also considered an alternate model in which any progenitor cell can dedifferentiate (Supplemental Material).
In this alternate model, dedifferentiation again had little effect for $\eps \lesssim 1/N_{sc}$.
Past that threshold the effect was substantial, because in this model dedifferentiation speeds up the time to emergence of two-mutation cells, because one-mutation cells fix much more quickly when they too can dedifferentiate (Supplemental Fig.~4D).
In addition, we considered the case in which the dedifferentiation rate is additionally weighted by the progenitor proliferation rate, and our results did not change qualitatively (Supplemental Material, Supplemental Fig.~4B).

Our analytical and numerical results suggest that, with intact homeostasis in the stem cell population and normal mutation rates, dedifferentiation plays a fairly minor role in speeding up the time to cancer initiation.
We thus turned to consider the case in which homeostasis is not strict.

\subsection*{Dedifferentiation with Variable Stem Cell Population Size}

In the previous section, we assumed that the stem cell population size was constant because homeostasis was maintained by all divisions being strictly asymmetric.
Consequently, dedifferentiated progenitor cells could only occupy newly created openings in the stem-cell niche created by a death event in the stem cell population.
Because homeostasis is likely maintained at the population level \cite{Lander}, with each stem cell division producing not strictly one stem cell but rather on average one stem cell, we next considered a model in which the stem cell population could stochastically fluctuate around a carrying capacity.
In this model, stem cell homeostasis was maintained by dynamically altering the probabilities of the three possible outcomes of a stem cell division: two stem cells, one stem and one progenitor cell, or two progenitor cells (Eq.~\ref{density-dependent GW 2}).
Two-mutation progenitor cells each had a probability per unit time of dedifferentiating, and dedifferentiated cells were simply added to the stem cell pool.
Thus in this model the total influx of dedifferentiated cells depended on the total number of two-mutation progenitor cells, not on the creation of openings in the stem cell niche.
(Note that in our previous model with constant stem cell population size the rate of dedifferentiation per reproduction event was denoted $\eps$.
To distinguish the present model, we denoted the progenitor dedifferentiation rate per cell per unit time as $\delta$.)
Again, we asked whether dedifferentiation substantially speeds the time to carcinogenesis.

Fig.~\ref{Figure: Model II variable generation_constant_growth}A and~\ref{Figure: Model II variable generation_constant_growth}B show typical results from this model for a moderate dedifferentiation rate $\delta$.
After a waiting time, the population of stems cells began to grow exponentially, because the influx of dedifferentiated two-mutation progenitor cells exceeded the capacity of stem-cell division homeostasis.
For larger dedifferentiation rates, the exponential growth rate is larger (Fig.~\ref{Figure: Model II variable generation_constant_growth}C and~\ref{Figure: Model II variable generation_constant_growth}D), and the distribution of progenitor ages can be distorted, with many young cells, as seen in Fig.~\ref{Figure: Model II variable generation_constant_growth}E and~\ref{Figure: Model II variable generation_constant_growth}F.

Exponential growth eventually occurs whenever the dedifferentiation rate exceeds a threshold $\delta_{crit}$.
Solving self-consistently for the influx of dedifferentiated cells and the growth rates of the stem and progenitor cell populations, we obtained an integral equation for the growth $k$
	\begin{equation}\label{eqn:full k}
	k= \alpha \delta \int_0^\infty e^{- k a} e^{r(a)}\, da-\frac{1-\eta}{T_{gen}},
	\end{equation}
which provides an excellent fit to the numerical simulations (Fig.~\ref{Figure: Model II variable generation_constant_growth} and~\ref{Figure:dedifferentiation_stats_Moran_variable}A,B). (For derivation details, see Supplemental Text.)
Setting this growth rate $k$ to zero, we found
\begin{equation}\label{eq:threshold1}
\delta_{crit} = \frac{1-\eta}{\alpha T_{gen} \int_0^\infty e^{r(a)}\, da}.
\end{equation}
Here $\eta$ is probability of asymmetric stem cell division (producing one stem and one progenitor cell), and $T_{gen}$ is the mean time between stem cell divisions. (Note that if $\eta = 1$, this model reduces to the Moran model with the population size monotonically increasing due to dedifferentiation.)  Lastly, $\alpha=2\alpha_{D,2}+\alpha_{A,2}$ in Eq.~\eqref{eq:threshold1} is the average number of progenitor offspring produced by a two-mutation stem cell.
Because $\alpha_{D,2}$ changes as the system attempts to maintain stem-cell homeostasis, $\alpha$ is actually a stochastic variable that depends on the stem cell population size. During exponential growth\ $\alpha \approx 2-\eta$, because the probability of symmetric divisions that give rise to two stem cells goes to zero, and all new stem cell growth comes from dedifferentiated progenitor cells.
In Eq.~\ref{eq:threshold1}, $r(a)$ is the growth rate of two-mutation progenitor cells as a function of age $a$, so $\int_0^\infty e^{r(a)}da$ is the number of progenitors produced by one two-mutation stem cell.
Increasing the amplification of mutant stem cells into progenitors increases the net dedifferentiation rate, lowering the threshold $\delta_{crit}$.
Because the threshold $\delta_{crit}$ depends on the age distribution of the two-mutation cells, for a given (small) rate of dedifferentiation $\delta$, evolving a mutant that proliferates faster (increasing $e^{r(a)}$) can destabilize a system in which the number of cancerous cells is stable  and take it into exponential growth regime.

The dependence of the critical dedifferentiation rate $\delta_{crit}$ on the growth-rate advantage $s$ of two-mutation progenitor cells and probability $\eta$ of asymmetric cell division is shown in Fig.~\ref{Figure:dedifferentiation_stats_Moran_variable}B.
The critical $\delta$ decreases rapidly as the selective advantage of two-mutation cells increases.
Increasing $\eta$ or $T_{gen}$ also lowers the critical dedifferentiation rate, because homeostasis is less effective when asymmetric stem cell divisions are less frequent. 		Note that the exponential growth rate $k$ does not depend on the mutation rate (Supplemental Fig.~5A), and although the critical $\delta$ given by \eqref{eq:threshold1} needed for exponential growth is a function of probability of asymmetric division $\eta$, the actual growth rate $k$ and the time to exponential growth is not significantly affected by changing $\eta$ (see Supplemental Fig.~5B).

For dedifferentiation rates $\delta$ below $\delta_{crit}$, two-mutation stem cells eventually fix in the population, but for $\delta > \delta_{crit}$, the stem cell population is likely to begin exponential growth before fixation of two-mutation stem cells.
Thus in Fig.~\ref{Figure:dedifferentiation_stats_Moran_variable}C and~\ref{Figure:dedifferentiation_stats_Moran_variable}D we report the time to carcinogenesis as the time for the two-mutation stem cell population to exceed $N_{sc}$, the nominal carrying capacity of the stem cell compartment.
In this case of stochastic stem cell homeostasis, dedifferentiation can dramatically shorten the time to carcinogenesis, even for low mutation rates $u$.
This is because the first two-mutation stem cell often arises not from direct mutation of a stem cell, but rather from dedifferentiation of a progenitor cell generated by mutations within the progenitor compartment (Fig.~\ref{Figure:dedifferentiation_stats_Moran_variable}E).
Although mutations in the progenitor compartment do not affect a large fraction of progenitors, because the number of progenitor cells is so large, the absolute number of two-mutation progenitor cells is non-negligible.
Thus even small rates of dedifferentiation can have dramatic effects. 
This is in contrast to the case of strict stem cell homeostasis, in which the absolute number of two-mutation progenitor cells was unimportant, because they needed an opening in the stem cell niche to successfully dedifferentiate.

Our results show that the case of stochastically controlled stem cell homeostasis is qualitatively different from the case of strict homeostasis.
If homeostasis is controlled at the population level (where decision between symmetric and asymmetric division are stochastic),  dedifferentiation can overwhelm it, leading to exponential growth of the stem cell population.
Moreover, if dedifferentiated cells do not depend on openings to colonize the stem cell niche, dedifferentiation can dramatically hasten the time to carcinogenesis, even for low mutation rates.

	\section*{Discussion}
	Progression to cancer is associated with expansion of the cancer stem cell (CSC) population, but the origin of these CSCs remains unclear.
	Although CSCs may arise directly from adult stem cells, they may also arise from somewhat differentiated cells that have dedifferentiated and acquired stem cell-like characteristics \cite{Passegue,Gupta-cell, Ischenko, Visvader}.
Stems cells replicate indefinitely, giving them a long time to accumulate the mutations that drive carcinogenesis, but the population of actively dividing stems cells ($N_{sc}$) is small.
Progenitor cells replicate only small number of times, but the population of progenitor cells is typically several orders of magnitude larger than the stem cell population. 
	Thus, as a population, progenitors undergo many more divisions, potentially letting some of these cells acquire mutations that enable them to dedifferentiate and drive carcinogenesis.
	Here, using mathematical modeling, we have shown that even a small rate of dedifferentiation may drastically shorten the time to cancer emergence.

	Recent studies suggest stem cell dynamics during homeostasis are governed by neutral competition and genetic drift~\cite{Calabrese_2004, lopez-garcia_science, klein_development}. 
	Traditionally, stem cells were thought to always undergo asymmetric division, always yielding a stem cell and a progenitor cell, resulting in a fixed stem cell population size.
	This scenario is represented by our first model for stem cell dynamics, based on the popular Moran model.
	It has been recently shown, however, that symmetric divisions also occur in adult stem cells and may be the predominant form of division \cite{SimonsClevers2011,Hu2013}.
	Moreover, cancer stem cells have been shown to undergo more symmetric divisions than normal stem cells~\cite{Cicalese_cell}.
	Little is known, however, about how the stem cell population size is regulated \cite{klein_development}.
	Hence, in our second model for stem cell dynamics, we made the simplifying assumption of an a priori carrying capacity $K_i$.
	We considered a density-dependent stochastic process, in which the degree of mean reversion is controlled through the probabilities of producing zero, one, or two stem cell offspring.
	In this model, the non-constant stem cell population size $S(t)$ tends to return to the carrying capacity $K_i$, because the mean number of stem cells produced per division is greater than one when $S(t)<K_i$ and less than one when $S(t)>K_i$.
	(Although the stem cell population size could, in principle, be maintained by regulating apoptosis rather than biasing division, previous modeling suggests that regulating division is more important for hematopoietic stem cells \cite{marciniak-czochra}.)

We compared the times to multiple mutation acquisition in our constant and variable stem cell population size models and found that without dedifferentiation both models yield similar results.
With dedifferentiation, however, we found that the two models differ substantially.
When the stem cell population size is constant, dedifferentiation simply acts like a selective advantage for mutant stem cells.
When the stem cell population size varies, however, dedifferentiation can additionally drive exponential growth of the stem cell population. 

If the stem cell population size $N_{sc}$ is constant, our results imply that stem cell dynamics in the coupled stem cell-progenitor system can be approximated by a population genetics model of the stem cells alone, as long as that model includes positive selection.
In this case, we found that the dedifferentiation rate $\varepsilon$ must exceed $1/N_{sc}$ to substantially shorten the time to cancer acquisition, similar to classical population genetics results that the selection coefficient must exceed the inverse population size to be effective.
	For the hematopoietic system, based on the literature we assumed that the number of actively dividing stem cells is $N_{sc}=100$, so the dedifferentation rate must be $\approx 0.01$ or higher to significantly shorten the time to cancer.

If the stem cell population size varies and is regulated by biasing division, we found two distinct regimes.
If the dedifferentiation rate is much less than a critical value, the initial two-mutation stem cell often arises from a normal stem cell, so the time to fixation of such a cell is similar to the case with constant population size.
If the dedifferentiation rate exceeds the critical value, however, the initial two-mutation stem cell often arises from a dedifferentiated progenitor cell, so the time to fixation is dramatically shorter than the case with constant population size.
Moreover, in this regime the stem cell population eventually grows exponentially, as dedifferentiating progenitor cells overwhelm stem cell homeostasis.
Note that the threshold between these two regimes is independent of the overall mutation rate, if stem and progenitor cell mutation rates are proportional.

The fact that mutants take a long time to reach an appreciable fraction of the stem cell population is not typically considered in the cancer modeling literature, which often makes an implicit assumption that a newly emerged mutant cell will not go extinct and will fix quickly. 
Our results show that, for high mutation rate, the time for a mutation to fix in the population is comparable to time for a successful mutant to first emerge, in accordance with classical results of Kimura and Ohta \cite{KimuraOhta1969}.
This is especially important if division events are rare and the population size is large.
Considering the time to some predetermined diagnosis threshold is similar to considering the time to fixation, because the time between a selected mutation becoming common and fixing is typically short~\cite{Patwa-review}. 
Hence, assuming elevated mutation rate (genetic instability) does not speed up time to carcinogenesis as much as is typically assumed, suggesting that some form of selection (potentially through dedifferentiation) is necessary.
Most tumors accumulate hundreds of mutations, but the number of necessary ``driver'' mutations depends on the type of cancer.
We considered $M=2$ mutations, because sequencing of acute myeloid leukemia genomes suggest that there are two driver mutations present \cite{Ley_nature}.
Moreover, recent findings on induced pluripotent stem cells also suggest $M=2$, as loss of both copies of the tumor suppressor protein p53 \cite{Spike_review} or the activation of two oncogenes \cite{Ischenko} may be necessaryfor dedifferentiation.
Disabling both copies of p53 improves the efficiency of reprogramming to a stem-like state and greatly enhances the production of induced pluripotent stem cells \cite{Kawamura, Hong2009}.
The loss of p53 also leads to the emergence of tumor cells bearing functional and molecular similarities to stem cells \cite{Spike_review, Zhao}.
Finally, inactivation of p53 changes the ratio of symmetric to asymmetric division in mammary stem cells, allowing the total stem cell population to escape homeostasis \cite{Cicalese_cell}.
(Inactivation of p53 will also lead to a higher effective mutation rate as cells with errors are permitted to continue in the cell cycle.)

Here we focus on the hematopoietic system, in which the stem cell compartment consists of $N_{sc} \approx 100$ active cells, and two mutations are necessary for carcinogenesis.
For some other cancers, such as colon cancer, the number of stem cells per compartment is much smaller, there are many compartments, and the number of necessary mutations is larger.
For high mutation rate, the mean time to fixation scales linearly with $N_{sc}$ (see Supplemental Fig.~4C.
So in cancers with small $N_{sc}$ two-mutation stem cells will fix much faster.
However, the need to accumulate more mutations will slow carcinogenesis.
We expect, however, that the qualitative effects of dedifferentiation will be similar to the hematopoietic system we analyzed.

Our model only considers actively dividing stem cells, which in the human hematopoietic system have been estimated to be roughly 100 \cite{Dingli_Cell_Cycle} out of 11,000-22,000 total stem cells \cite{abkowitz_blood}. 
A more complete model would consider both the active and quiescent stem cell populations.
Transitions between these states may be influenced by the progenitor population size, potentially acting as a negative feedback and regulating the proliferation of cancer stem cells. 
In our models, cancerous cells take over the stem cell population, but the ratio of cancer progenitor cells to cancer stem cells is fixed by the progenitor growth process.
Even when dedifferentiation drives exponential growth of the stem cells, it is their absolute number that increases, not their proportion in the population.
This is in concordance with some \textit{in vitro} studies, which suggest a fixed proportion of CSCs in a tumor \cite{Gupta-cell}.
	
Many theoretical models find that in order to accumulate multiple mutations on a reasonable time scale, the onset of elevated mutation rate (i.e., genetic instability) should be an early event in tumorigenesis (reviewed in \cite{attolini-Michor, Beckman_Loeb}).
The importance of genetic instability, however, depends on assumptions about symmetric self-renewal and differentiation of stem and progenitor cells.
In particular, mutations that alter stem cell division or make committed progenitors somewhat immortal may also lead to an early onset of cancer, diminishing the impact of genetic instability \cite{ashkenazi-neoplasia}. 
Similarly, our results show that different assumptions about how dedifferentiation occurs (frequency-dependent reproduction versus absolute numbers of dedifferentiating cells) dramatically alter time to carcinogenesis.
	
Like other mathematical models, our model suggests that eradication of cancer is dependent on eradication of cancer stem cells~\cite{Leder_PlosOne,Sehl_CancerResearch, Sehl_MathBiosci}.
The potential for progenitor cells to dedifferentiate and repopulate the stem cell compartment, however, may complicate successful treatment.
Our work suggests that further progress in understanding initiation and treatment of cancer requires a more detailed understanding dedifferentiation and of stem cell homeostasis.

	\section*{Acknowledgments}

	\bibliography{Cancer}

\begin{thebibliography}{10}
\providecommand{\url}[1]{\texttt{#1}}
\providecommand{\urlprefix}{URL }
\expandafter\ifx\csname urlstyle\endcsname\relax
  \providecommand{\doi}[1]{doi:\discretionary{}{}{}#1}\else
  \providecommand{\doi}{doi:\discretionary{}{}{}\begingroup
  \urlstyle{rm}\Url}\fi
\providecommand{\bibAnnoteFile}[1]{%
  \IfFileExists{#1}{\begin{quotation}\noindent\textsc{Key:} #1\\
  \textsc{Annotation:}\ \input{#1}\end{quotation}}{}}
\providecommand{\bibAnnote}[2]{%
  \begin{quotation}\noindent\textsc{Key:} #1\\
  \textsc{Annotation:}\ #2\end{quotation}}
\providecommand{\eprint}[2][]{\url{#2}}

\bibitem{reya_nature_review}
Reya T, Morrison SJ, Clarke MF, Weissman IL (2001) {{S}tem cells, cancer, and
  cancer stem cells}.
\newblock Nature 414: 105--111.
\bibAnnoteFile{reya_nature_review}

\bibitem{Driessens_Nature2012}
Driessens G, Beck B, Caauwe A, Simons BD, Blanpain C (2012) {{D}efining the
  mode of tumour growth by clonal analysis}.
\newblock Nature 488: 527--530.
\bibAnnoteFile{Driessens_Nature2012}

\bibitem{medema}
Medema JP (2013) {{C}ancer stem cells: the challenges ahead}.
\newblock Nat Cell Biol 15: 338--344.
\bibAnnoteFile{medema}

\bibitem{attolini-Michor}
Attolini CS, Michor F (2009) {{E}volutionary theory of cancer}.
\newblock Ann N Y Acad Sci 1168: 23--51.
\bibAnnoteFile{attolini-Michor}

\bibitem{Michor_Nowak_2006}
Michor F, Iwasa Y, Nowak MA (2006) {{T}he age incidence of chronic myeloid
  leukemia can be explained by a one-mutation model}.
\newblock Proc Natl Acad Sci USA 103: 14931--14934.
\bibAnnoteFile{Michor_Nowak_2006}

\bibitem{Beerenwinkel}
Beerenwinkel N, Antal T, Dingli D, Traulsen A, Kinzler KW, et~al. (2007)
  {{G}enetic progression and the waiting time to cancer}.
\newblock PLoS Comput Biol 3: e225.
\bibAnnoteFile{Beerenwinkel}

\bibitem{Beckman_Loeb}
Beckman RA, Loeb LA (2005) {{G}enetic instability in cancer: theory and
  experiment}.
\newblock Semin Cancer Biol 15: 423--435.
\bibAnnoteFile{Beckman_Loeb}

\bibitem{Abdel-Rahman}
Abdel-Rahman WM (2008) {{G}enomic instability and carcinogenesis: an update}.
\newblock Curr Genomics 9: 535--541.
\bibAnnoteFile{Abdel-Rahman}

\bibitem{Klein2006}
Klein CA (2006) {{R}andom mutations, selected mutations: {A} {P}{I}{N} opens
  the door to new genetic landscapes}.
\newblock Proc Natl Acad Sci USA 103: 18033--18034.
\bibAnnoteFile{Klein2006}

\bibitem{Calabrese_2004}
Calabrese P, Tavare S, Shibata D (2004) {{P}retumor progression: clonal
  evolution of human stem cell populations}.
\newblock Am J Pathol 164: 1337--1346.
\bibAnnoteFile{Calabrese_2004}

\bibitem{Jones_PNAS2008}
Jones S, Chen WD, Parmigiani G, Diehl F, Beerenwinkel N, et~al. (2008)
  {{C}omparative lesion sequencing provides insights into tumor evolution}.
\newblock Proc Natl Acad Sci USA 105: 4283--4288.
\bibAnnoteFile{Jones_PNAS2008}

\bibitem{wang_PNAS}
Wang TL, Rago C, Silliman N, Ptak J, Markowitz S, et~al. (2002) {{P}revalence
  of somatic alterations in the colorectal cancer cell genome}.
\newblock Proc Natl Acad Sci USA 99: 3076--3080.
\bibAnnoteFile{wang_PNAS}

\bibitem{Visvader}
Visvader JE (2011) {{C}ells of origin in cancer}.
\newblock Nature 469: 314--322.
\bibAnnoteFile{Visvader}

\bibitem{Passegue}
Passegue E, Jamieson CH, Ailles LE, Weissman IL (2003) {{N}ormal and leukemic
  hematopoiesis: are leukemias a stem cell disorder or a reacquisition of stem
  cell characteristics?}
\newblock Proc Natl Acad Sci USA 100 Suppl 1: 11842--11849.
\bibAnnoteFile{Passegue}

\bibitem{Krivtsov}
Krivtsov AV, Twomey D, Feng Z, Stubbs MC, Wang Y, et~al. (2006)
  {{T}ransformation from committed progenitor to leukaemia stem cell initiated
  by {M}{L}{L}-{A}{F}9}.
\newblock Nature 442: 818--822.
\bibAnnoteFile{Krivtsov}

\bibitem{Goardon}
Goardon N, Marchi E, Atzberger A, Quek L, Schuh A, et~al. (2011) {{C}oexistence
  of {L}{M}{P}{P}-like and {G}{M}{P}-like leukemia stem cells in acute myeloid
  leukemia}.
\newblock Cancer Cell 19: 138--152.
\bibAnnoteFile{Goardon}

\bibitem{Zhao}
Zhao Z, Zuber J, Diaz-Flores E, Lintault L, Kogan SC, et~al. (2010) {p53 loss
  promotes acute myeloid leukemia by enabling aberrant self-renewal}.
\newblock Genes Dev 24: 1389--1402.
\bibAnnoteFile{Zhao}

\bibitem{Dingli_BMC_Biol}
Dingli D, Pacheco JM (2011) {{S}tochastic dynamics and the evolution of
  mutations in stem cells}.
\newblock BMC Biol 9: 41.
\bibAnnoteFile{Dingli_BMC_Biol}

\bibitem{Jamieson}
Jamieson CH, Ailles LE, Dylla SJ, Muijtjens M, Jones C, et~al. (2004)
  {{G}ranulocyte-macrophage progenitors as candidate leukemic stem cells in
  blast-crisis {C}{M}{L}}.
\newblock N Engl J Med 351: 657--667.
\bibAnnoteFile{Jamieson}

\bibitem{Haeno_PNAS}
Haeno H, Levine RL, Gilliland DG, Michor F (2009) {{A} progenitor cell origin
  of myeloid malignancies}.
\newblock Proc Natl Acad Sci USA 106: 16616--16621.
\bibAnnoteFile{Haeno_PNAS}

\bibitem{FrankIwasaNowak2003}
Frank SA, Iwasa Y, Nowak MA (2003) {{P}atterns of cell division and the risk of
  cancer}.
\newblock Genetics 163: 1527--1532.
\bibAnnoteFile{FrankIwasaNowak2003}

\bibitem{Komarova-review}
Komarova NL (2005) {{C}ancer, aging and the optimal tissue design}.
\newblock Semin Cancer Biol 15: 494--505.
\bibAnnoteFile{Komarova-review}

\bibitem{Wodarz_book}
Wodarz D, Komarova N (2005) Computational Biology of Cancer: Lecture Notes and
  Mathematical Modeling.
\newblock World Scientific Publishing Company.
\bibAnnoteFile{Wodarz_book}

\bibitem{klein_development}
Klein AM, Simons BD (2011) {{U}niversal patterns of stem cell fate in cycling
  adult tissues}.
\newblock Development 138: 3103--3111.
\bibAnnoteFile{klein_development}

\bibitem{lopez-garcia_science}
Lopez-Garcia C, Klein AM, Simons BD, Winton DJ (2010) {{I}ntestinal stem cell
  replacement follows a pattern of neutral drift}.
\newblock Science 330: 822--825.
\bibAnnoteFile{lopez-garcia_science}

\bibitem{snippert_cell}
Snippert HJ, van~der Flier LG, Sato T, van Es JH, van~den Born M, et~al. (2010)
  {{I}ntestinal crypt homeostasis results from neutral competition between
  symmetrically dividing {L}gr5 stem cells}.
\newblock Cell 143: 134--144.
\bibAnnoteFile{snippert_cell}

\bibitem{Dingli_Cell_Cycle}
Dingli D, Traulsen A, Pacheco JM (2007) {{S}tochastic dynamics of hematopoietic
  tumor stem cells}.
\newblock Cell Cycle 6: 461--466.
\bibAnnoteFile{Dingli_Cell_Cycle}

\bibitem{Iwasa2004}
Iwasa Y, Michor F, Nowak MA (2004) {{S}tochastic tunnels in evolutionary
  dynamics}.
\newblock Genetics 166: 1571--1579.
\bibAnnoteFile{Iwasa2004}

\bibitem{Durrett2009}
Durrett R, Schmidt D, Schweinsberg J (2009) {A waiting time problem arising
  from the study of multi-stage carcinogenesis}.
\newblock Ann Appl Prob 19: 676--718.
\bibAnnoteFile{Durrett2009}

\bibitem{Dingli_PLOS_CB}
Dingli D, Traulsen A, Michor F, Michor F (2007) {({A})symmetric stem cell
  replication and cancer}.
\newblock PLoS Comput Biol 3: e53.
\bibAnnoteFile{Dingli_PLOS_CB}

\bibitem{Dingli_PhysRevE}
Dingli D, Pacheco JM, Traulsen A (2008) {{M}ultiple mutant clones in blood
  rarely coexist}.
\newblock Phys Rev E Stat Nonlin Soft Matter Phys 77: 021915.
\bibAnnoteFile{Dingli_PhysRevE}

\bibitem{Foo2011}
Foo J, Leder K, Michor F (2011) {{S}tochastic dynamics of cancer initiation}.
\newblock Phys Biol 8: 015002.
\bibAnnoteFile{Foo2011}

\bibitem{Thalhauser}
Thalhauser CJ, Lowengrub JS, Stupack D, Komarova NL (2010) {{S}election in
  spatial stochastic models of cancer: migration as a key modulator of
  fitness}.
\newblock Biol Direct 5: 21.
\bibAnnoteFile{Thalhauser}

\bibitem{yatabe_PNAS}
Yatabe Y, Tavare S, Shibata D (2001) {{I}nvestigating stem cells in human colon
  by using methylation patterns}.
\newblock Proc Natl Acad Sci USA 98: 10839--10844.
\bibAnnoteFile{yatabe_PNAS}

\bibitem{Bozic}
Bozic I, Antal T, Ohtsuki H, Carter H, Kim D, et~al. (2010) {{A}ccumulation of
  driver and passenger mutations during tumor progression}.
\newblock Proc Natl Acad Sci USA 107: 18545--18550.
\bibAnnoteFile{Bozic}

\bibitem{Durrett2011}
Durrett R, Foo J, Leder K, Mayberry J, Michor F (2011) {{I}ntratumor
  heterogeneity in evolutionary models of tumor progression}.
\newblock Genetics 188: 461--477.
\bibAnnoteFile{Durrett2011}

\bibitem{Haeno_genetics}
Haeno H, Iwasa Y, Michor F (2007) {{T}he evolution of two mutations during
  clonal expansion}.
\newblock Genetics 177: 2209--2221.
\bibAnnoteFile{Haeno_genetics}

\bibitem{Tomasetti2012}
Tomasetti C (2012) {{O}n the probability of random genetic mutations for
  various types of tumor growth}.
\newblock Bull Math Biol 74: 1379--1395.
\bibAnnoteFile{Tomasetti2012}

\bibitem{spencer_2006}
Spencer SL, Gerety RA, Pienta KJ, Forrest S (2006) {{M}odeling somatic
  evolution in tumorigenesis}.
\newblock PLoS Comput Biol 2: e108.
\bibAnnoteFile{spencer_2006}

\bibitem{ashkenazi-neoplasia}
Ashkenazi R, Gentry SN, Jackson TL (2008) {{P}athways to
  tumorigenesis--modeling mutation acquisition in stem cells and their
  progeny}.
\newblock Neoplasia 10: 1170--1182.
\bibAnnoteFile{ashkenazi-neoplasia}

\bibitem{marciniak-czochra}
Marciniak-Czochra A, Stiehl T, Ho AD, Jager W, Wagner W (2009) {{M}odeling of
  asymmetric cell division in hematopoietic stem cells--regulation of
  self-renewal is essential for efficient repopulation}.
\newblock Stem Cells Dev 18: 377--385.
\bibAnnoteFile{marciniak-czochra}

\bibitem{Johnston_PNAS}
Johnston MD, Edwards CM, Bodmer WF, Maini PK, Chapman SJ (2007) {{M}athematical
  modeling of cell population dynamics in the colonic crypt and in colorectal
  cancer}.
\newblock Proc Natl Acad Sci USA 104: 4008--4013.
\bibAnnoteFile{Johnston_PNAS}

\bibitem{mahaffy_jtb}
Mahaffy JM, Belair J, Mackey MC (1998) {{H}ematopoietic model with moving
  boundary condition and state dependent delay: applications in
  erythropoiesis}.
\newblock J Theor Biol 190: 135--146.
\bibAnnoteFile{mahaffy_jtb}

\bibitem{marciniak-czochra_siam}
Doumic M, Marciniak-Czochra A, Perthame B, Zubelli J (2011) {A Structured
  Population Model of Cell Differentiation }.
\newblock SIAM J Appl Math 71: 1918Ð-1940.
\bibAnnoteFile{marciniak-czochra_siam}

\bibitem{Tomasetti2010}
Tomasetti C, Levy D (2010) {{R}ole of symmetric and asymmetric division of stem
  cells in developing drug resistance}.
\newblock Proc Natl Acad Sci USA 107: 16766--16771.
\bibAnnoteFile{Tomasetti2010}

\bibitem{Ewens_book}
Ewens W (2004) Mathematical Population Genetics.
\newblock Springer.
\bibAnnoteFile{Ewens_book}

\bibitem{Dingli_2010}
Dingli D, Traulsen A, Lenaerts T, Pacheco JM (2010) {{E}volutionary dynamics of
  chronic myeloid leukemia}.
\newblock Genes Cancer 1: 309--315.
\bibAnnoteFile{Dingli_2010}

\bibitem{Dingli_PNAS}
Dingli D, Luzzatto L, Pacheco JM (2008) {{N}eutral evolution in paroxysmal
  nocturnal hemoglobinuria}.
\newblock Proc Natl Acad Sci USA 105: 18496--18500.
\bibAnnoteFile{Dingli_PNAS}

\bibitem{Ley_nature}
Ley TJ, Mardis ER, Ding L, Fulton B, McLellan MD, et~al. (2008) {{D}{N}{A}
  sequencing of a cytogenetically normal acute myeloid leukaemia genome}.
\newblock Nature 456: 66--72.
\bibAnnoteFile{Ley_nature}

\bibitem{Crow_book}
Crow J, Kimura M (2011) An introduction to population genetics theory.
\newblock Blackburn Press.
\bibAnnoteFile{Crow_book}

\bibitem{KimuraOhta1969}
Kimura M, Ohta T (1969) {The average number of generations until fixation of a
  mutant gene in a finite population}.
\newblock Genetics 61: 763--771.
\bibAnnoteFile{KimuraOhta1969}

\bibitem{Allen_book}
Allen LJ (2011) An introduction to stochastic processes with applications to
  biology.
\newblock CRC Press.
\bibAnnoteFile{Allen_book}

\bibitem{Parsons_genetics}
Parsons TL, Quince C, Plotkin JB (2010) {{S}ome consequences of demographic
  stochasticity in population genetics}.
\newblock Genetics 185: 1345--1354.
\bibAnnoteFile{Parsons_genetics}

\bibitem{abkowitz_blood}
Abkowitz JL, Catlin SN, McCallie MT, Guttorp P (2002) {{E}vidence that the
  number of hematopoietic stem cells per animal is conserved in mammals}.
\newblock Blood 100: 2665--2667.
\bibAnnoteFile{abkowitz_blood}

\bibitem{Nash}
Nash R, Storb R, Neiman P (1988) {{P}olyclonal reconstitution of human marrow
  after allogeneic bone marrow transplantation}.
\newblock Blood 72: 2031--2037.
\bibAnnoteFile{Nash}

\bibitem{Catlin}
Catlin SN, Busque L, Gale RE, Guttorp P, Abkowitz JL (2011) {{T}he replication
  rate of human hematopoietic stem cells in vivo}.
\newblock Blood 117: 4460--4466.
\bibAnnoteFile{Catlin}

\bibitem{Traulsen_bioessays}
Traulsen A, Pacheco JM, Luzzatto L, Dingli D (2010) {{S}omatic mutations and
  the hierarchy of hematopoiesis}.
\newblock Bioessays 32: 1003--1008.
\bibAnnoteFile{Traulsen_bioessays}

\bibitem{Nowak_book}
Nowak M (2006) Evolutionary Dynamics.
\newblock Harvard University Press.
\bibAnnoteFile{Nowak_book}

\bibitem{araten}
Araten DJ, Golde DW, Zhang RH, Thaler HT, Gargiulo L, et~al. (2005) {{A}
  quantitative measurement of the human somatic mutation rate}.
\newblock Cancer Res 65: 8111--8117.
\bibAnnoteFile{araten}

\bibitem{Parsons_2008}
Parsons TL, Quince C, Plotkin JB (2008) {{A}bsorption and fixation times for
  neutral and quasi-neutral populations with density dependence}.
\newblock Theor Popul Biol 74: 302--310.
\bibAnnoteFile{Parsons_2008}

\bibitem{Lander}
Lander AD (2011) {{T}he individuality of stem cells}.
\newblock BMC Biol 9: 40.
\bibAnnoteFile{Lander}

\bibitem{Gupta-cell}
Gupta PB, Fillmore CM, Jiang G, Shapira SD, Tao K, et~al. (2011) {{S}tochastic
  state transitions give rise to phenotypic equilibrium in populations of
  cancer cells}.
\newblock Cell 146: 633--644.
\bibAnnoteFile{Gupta-cell}

\bibitem{Ischenko}
Ischenko I, Zhi J, Moll UM, Nemajerova A, Petrenko O (2013) {{D}irect
  reprogramming by oncogenic {R}as and {M}yc}.
\newblock Proc Natl Acad Sci USA 110: 3937--3942.
\bibAnnoteFile{Ischenko}

\bibitem{SimonsClevers2011}
Simons BD, Clevers H (2011) {{S}trategies for homeostatic stem cell
  self-renewal in adult tissues}.
\newblock Cell 145: 851--862.
\bibAnnoteFile{SimonsClevers2011}

\bibitem{Hu2013}
Hu Z, Fu YX, Greenberg AJ, Wu CI, Zhai W (2013) {{A}ge-dependent transition
  from cell-level to population-level control in murine intestinal homeostasis
  revealed by coalescence analysis}.
\newblock PLoS Genet 9: e1003326.
\bibAnnoteFile{Hu2013}

\bibitem{Cicalese_cell}
Cicalese A, Bonizzi G, Pasi CE, Faretta M, Ronzoni S, et~al. (2009) {{T}he
  tumor suppressor p53 regulates polarity of self-renewing divisions in mammary
  stem cells}.
\newblock Cell 138: 1083--1095.
\bibAnnoteFile{Cicalese_cell}

\bibitem{Patwa-review}
Patwa Z, Wahl LM (2008) {{T}he fixation probability of beneficial mutations}.
\newblock J R Soc Interface 5: 1279--1289.
\bibAnnoteFile{Patwa-review}

\bibitem{Spike_review}
Spike BT, Wahl GM (2011) {p53, {S}tem {C}ells, and {R}eprogramming: {T}umor
  {S}uppression beyond {G}uarding the {G}enome}.
\newblock Genes Cancer 2: 404--419.
\bibAnnoteFile{Spike_review}

\bibitem{Kawamura}
Kawamura T, Suzuki J, Wang YV, Menendez S, Morera LB, et~al. (2009) {{L}inking
  the p53 tumour suppressor pathway to somatic cell reprogramming}.
\newblock Nature 460: 1140--1144.
\bibAnnoteFile{Kawamura}

\bibitem{Hong2009}
Hong H, Takahashi K, Ichisaka T, Aoi T, Kanagawa O, et~al. (2009)
  {{S}uppression of induced pluripotent stem cell generation by the p53-p21
  pathway}.
\newblock Nature 460: 1132--1135.
\bibAnnoteFile{Hong2009}

\bibitem{Leder_PlosOne}
Leder K, Holland EC, Michor F (2010) {{T}he therapeutic implications of
  plasticity of the cancer stem cell phenotype}.
\newblock PLoS ONE 5: e14366.
\bibAnnoteFile{Leder_PlosOne}

\bibitem{Sehl_CancerResearch}
Sehl ME, Sinsheimer JS, Zhou H, Lange KL (2009) {{D}ifferential destruction of
  stem cells: implications for targeted cancer stem cell therapy}.
\newblock Cancer Res 69: 9481--9489.
\bibAnnoteFile{Sehl_CancerResearch}

\bibitem{Sehl_MathBiosci}
Sehl M, Zhou H, Sinsheimer JS, Lange KL (2011) {{E}xtinction models for cancer
  stem cell therapy}.
\newblock Math Biosci 234: 132--146.
\bibAnnoteFile{Sehl_MathBiosci}

\bibitem{lutscher}
Lutscher F, McCauley E, Lewis M (2007) {Spatial patterns and coexistence
  mechanisms in rivers. }.
\newblock Theor Pop Biol 71: 267Ð-277.
\bibAnnoteFile{lutscher}

\bibitem{Murray_book}
Murray J (2003) Mathematical Biology.
\newblock Springer.
\bibAnnoteFile{Murray_book}

\end{thebibliography}

	\section*{Figure Legends}
	\begin{figure}[!ht]
	\begin{centering}
	\includegraphics[width=12.35cm]{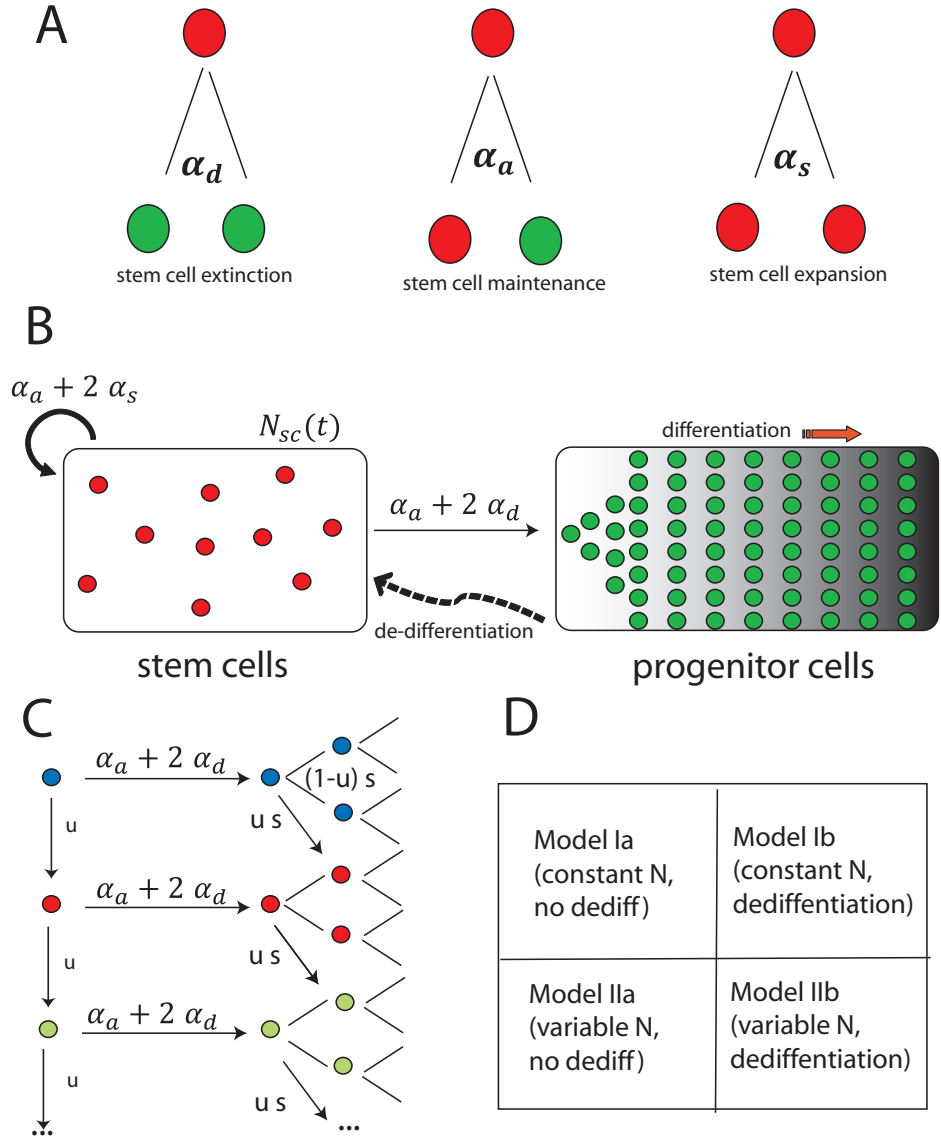}
	\par\end{centering}
	\caption{\textbf{Schematic representation of our model.} (A) Stochastic model for stem cell division. A stem cell can produce zero, one, or two stem cells with probabilities $\alpha_D$, $\alpha_A$, and $\alpha_S$, respectively.The mean number of offering is given by $m=\alpha_A+2 \alpha_S$. (B) Stem cells serve as an input to the proliferating progenitor population, and the progenitor population feeds back to the stem cell pool via dedifferentiation. (C) Mutation occurs with rate $u$ during division and can affect cells both in the stem cell and progenitor pools. Blue circles represent wild-type cells, red circles cells with one mutation, and green circles cells with two mutations. (D) The sequence of models explored in this paper.
	}\label{Figure1:schematic}
	\end{figure}
	
	\begin{figure}[!ht]
	\begin{centering}
	\includegraphics[width=17.35cm]{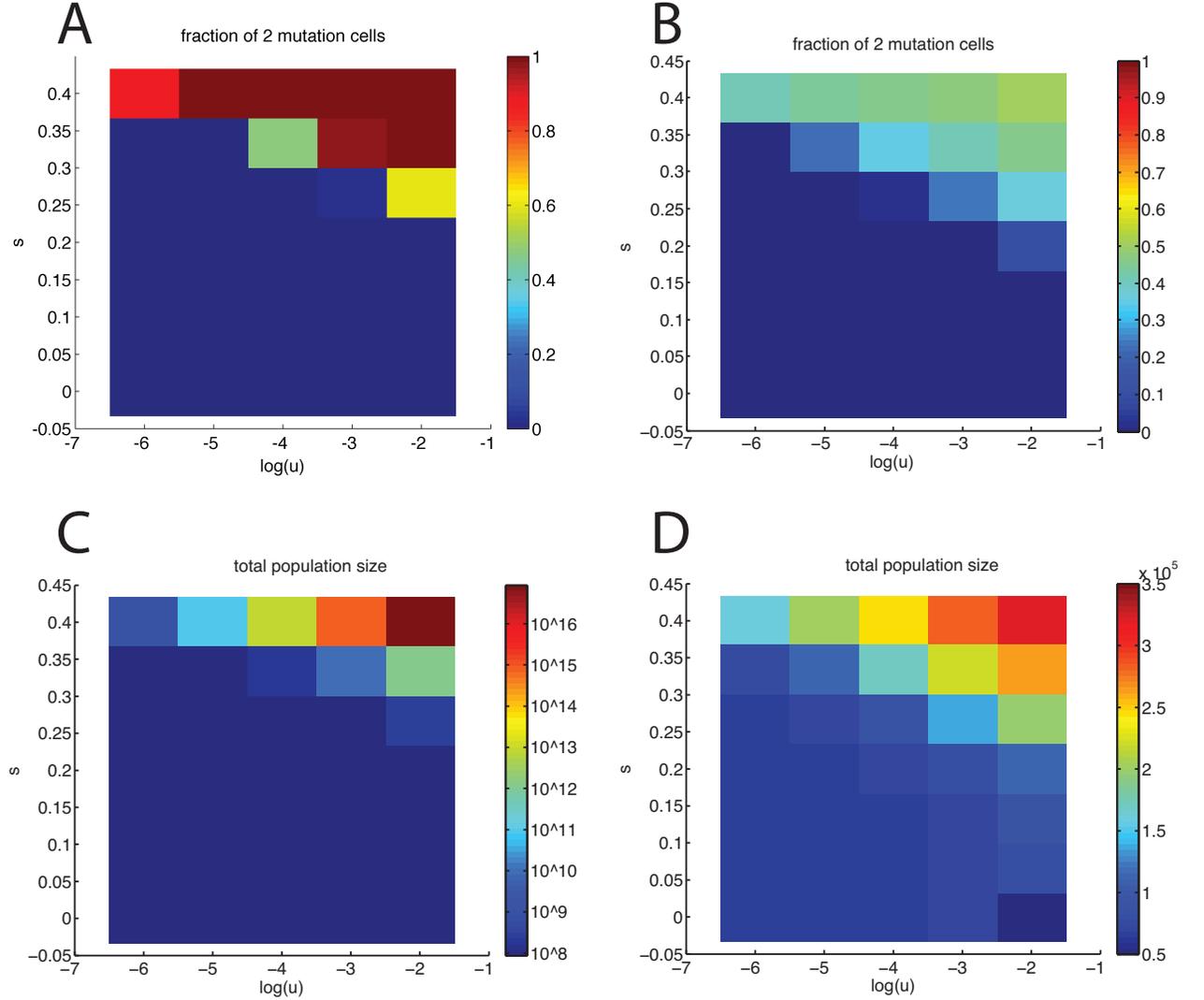}
	\par\end{centering}
	\caption{
	\textbf{Steady-state progenitor distributions in the absence of stem cell mutation.}
	Top: Fraction of two-mutation cells as a function of mutation rate $u$ and proliferative advantage $s$ for (a) unlimited growth, (b) logistic growth for each subpopulation.
	Bottom: Corresponding total population sizes.
		Birth/death rates of progenitor cells are given by \eqref {equations:birth_death} with constant death rate $\mu=1$ and sigmoidal birth rate with maximal growth rate $b_0=2$, $b_i=(1+s)b_{i-1}$ for $i=1,2$.
	In (b) the carrying capacity used is $N_1=200 N_{sc}, \, N_2=250 N_{sc},\, N_3=300 N_{sc}.$ Other parameters are as in Table 1.
	For two-mutation cells to reach appreciable levels in this scenario, both the mutation rate and proliferative advantage must be unreasonably large.}\label{Figure:competition plots}
	\end{figure}
	
	\begin{figure}[!ht]
	\begin{centering}
	\includegraphics[width=17.35cm]{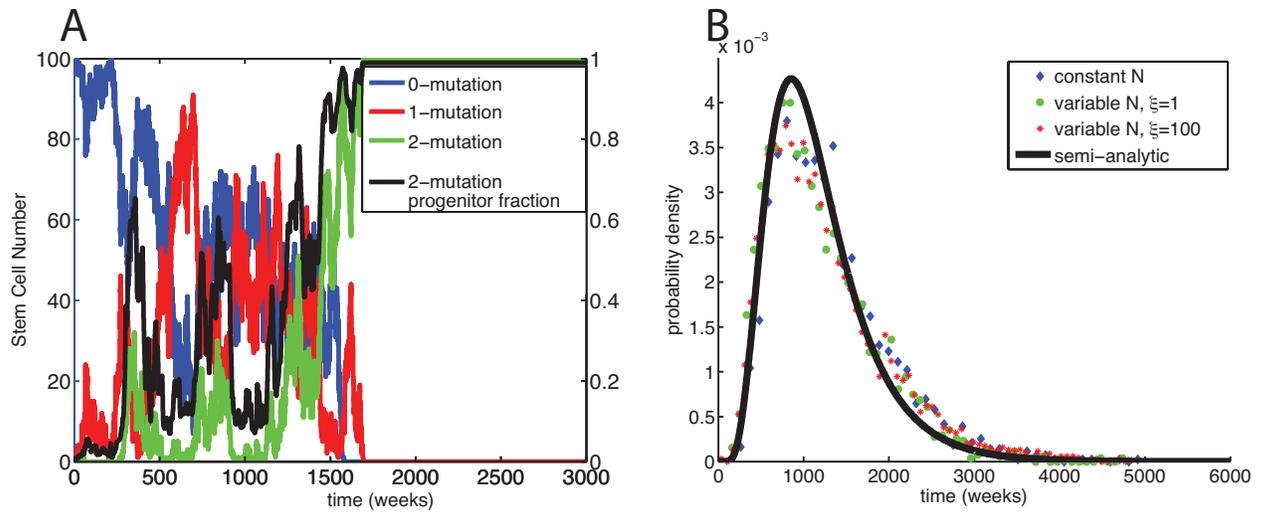}
	\par\end{centering}
\caption{ \textbf{Times to fixation without dedifferentiation.}
(A) Typical simulation trajectory for constant stem cell population size and mutation rate $u=0.1$. The numbers of zero-, one-, and two-mutation stem cells are shown in blue, red, and green, respectively. The proportion of two-mutation cells in the progenitor population is shown in black.
(B) Times to fixation for constant and variable stem cell population size models. Histogram of waiting times to fixation of two-mutation cells for constant (blue) and variable stem cell population size with high fluctuations (green, $\xi=1$) and low fluctuations (red, $\xi=100$). 
The semi-analytic distribution of waiting times calculated from Eq.~\eqref{eq:ultimate_prob_fixation}) is shown in black. In both panels the mutation rate $u = 0.01$. 
}
\label{Figure:Moran_waiting_time}
\end{figure}
	
\begin{figure}[!ht]
	\begin{centering}
	\includegraphics[width=17.35cm]{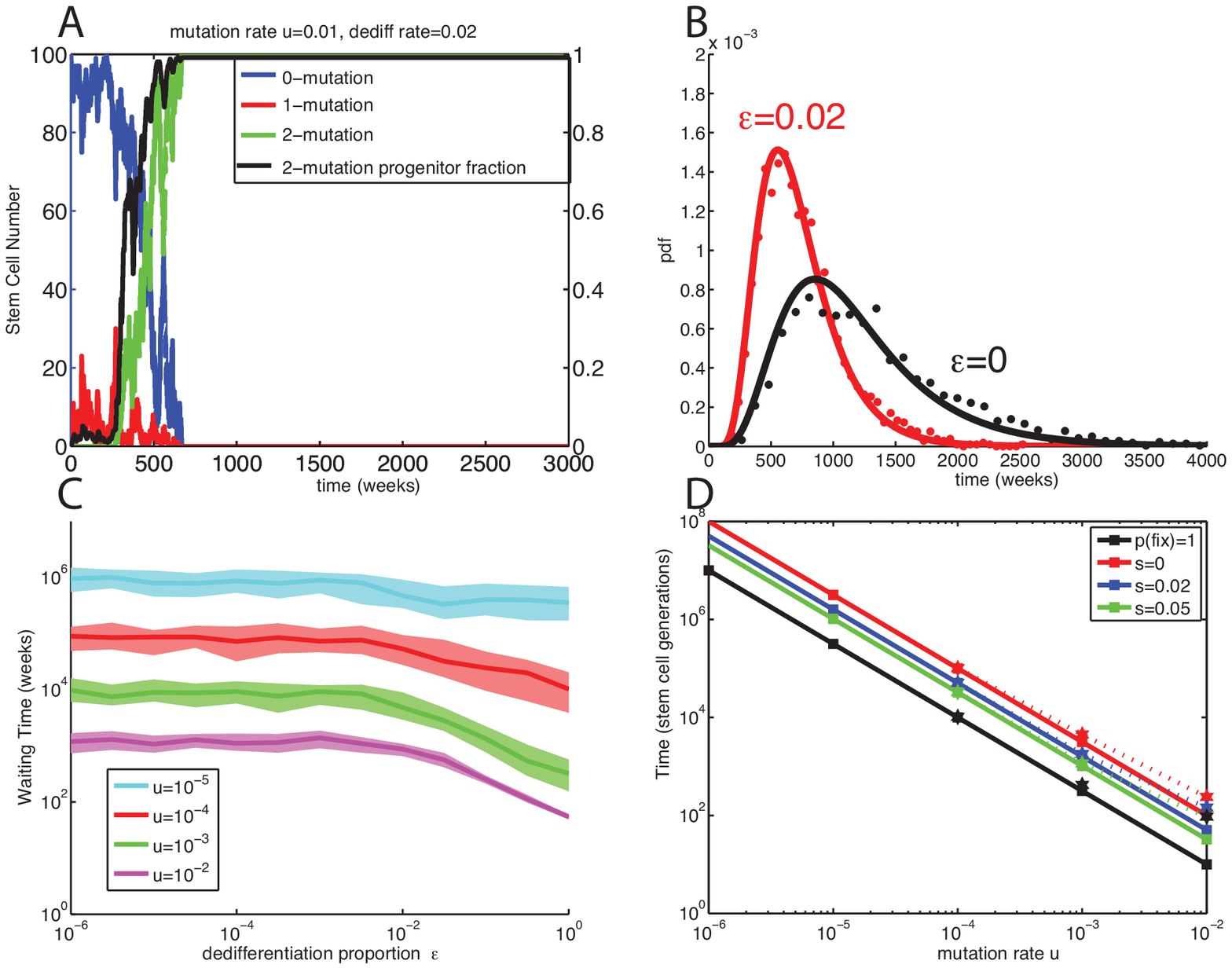}
	\par\end{centering}
\caption{\textbf{Times to fixation with dedifferentiation for constant stem cell population size}
 (A) Typical simulation trajectory with dedifferentiation ($\eps=0.02$) for the same random number seed as Fig 3A. Blue: zero-mutation stem cells, Red: one-mutation stem cells, Green: two-mutation stem cells. Black: proportion of two-mutation cells in the progenitor population.
(B) Distributions of times to fixation of two-mutation cells under strict stem cell homeostasis.
Normalized histograms (dots) and analytical approximations (solid lines) are shown for $u=0.01$ for
zero dedifferentiation (red; $\varepsilon=0$) and non-zero dedifferentiation (black; $\varepsilon=0.02$).
(C) 
Median times to fixation of two-mutation cells (solid lines) and inter-quantile ranges (shaded regions) versus dedifferentiation rate $\eps$ and mutation rate $u$.
(D) Mean times to emergence of a successful two-mutation stem cell (solid lines, Eq.~\eqref{eq:Durrett_modified}) and fixation of such cells (dotted lines, Eq.~\eqref{eq:ultimate_prob_fixation}) in Moran models with selection coefficient shown. Black curve indicates first appearance of two-mutant cell.
}\label{Figure:dedifferentiation_stats_Moran_generations}
\end{figure}

	\begin{figure}[htbp]
	\begin{centering}
	\includegraphics[width=17.35cm]{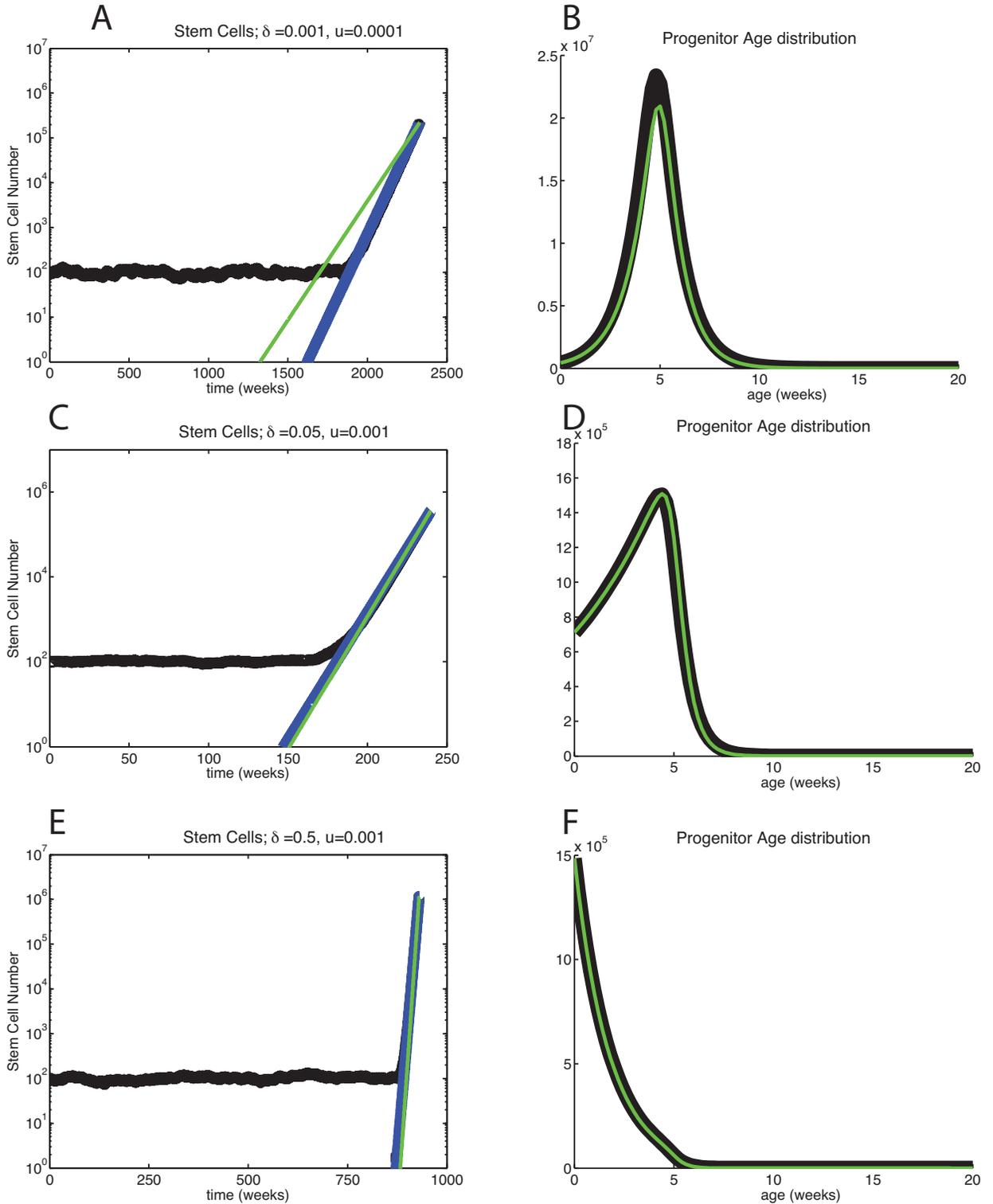}
	\par\end{centering}
	\caption{ \textbf{Exponential growth given varying stem cell population size and dedifferentiation}
	Total number of stem cells (A, C, E) and corresponding final progenitor age distributions (B, D, F) are shown in black.
	Also shown are best exponential fits of the growth rate (blue) and our semi-analytic solution given by \eqref{eqn:full k} (green).
	In all panels the probability of asymmetric stem cell division $\eta=0$ and the mean reversion parameter is $\xi=1$.
	}
	\label{Figure: Model II variable generation_constant_growth}
	\end{figure}
	
	\begin{figure}[htbp]
	\begin{centering}
	\includegraphics[width=17.35cm]{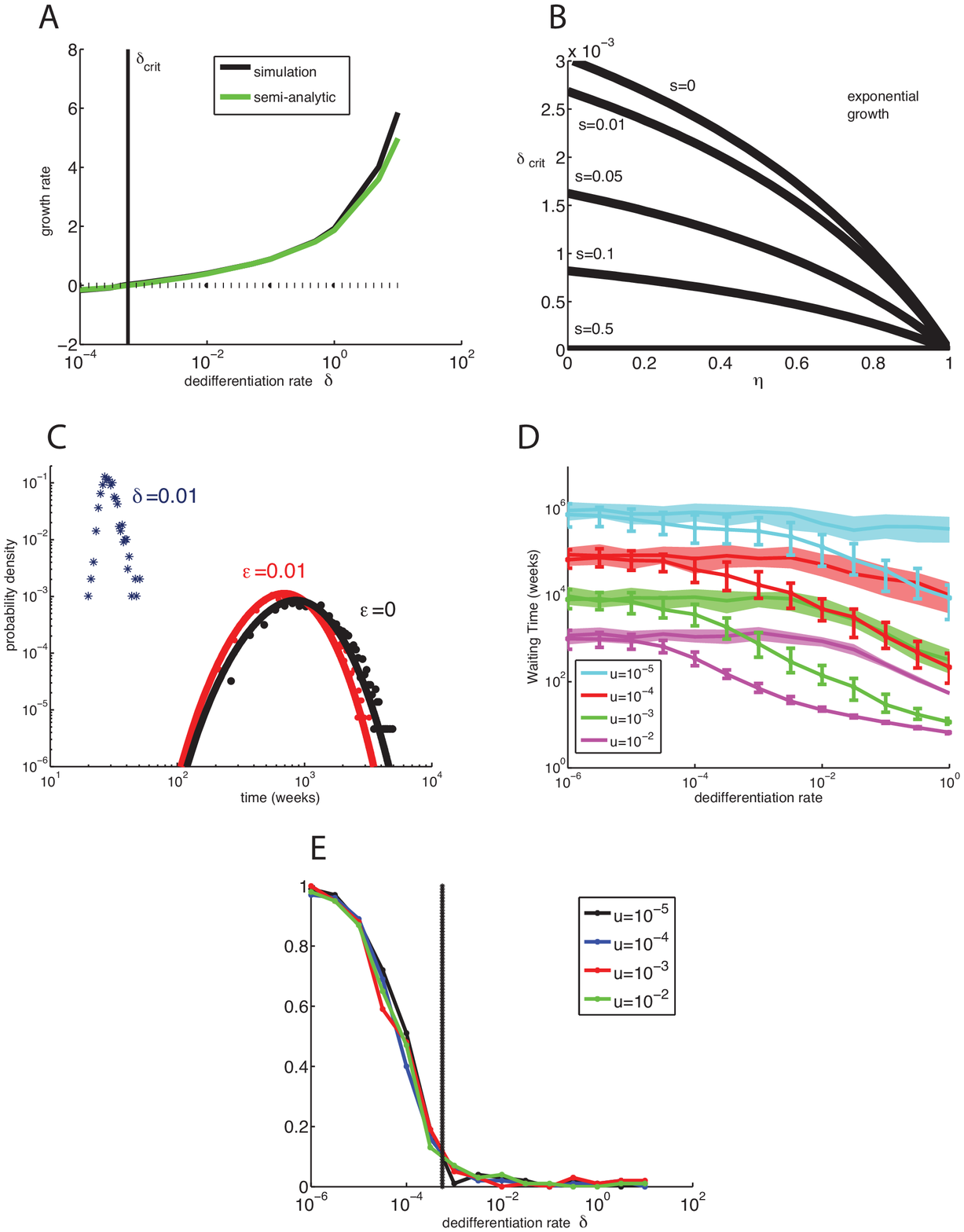}
	\par\end{centering}
\end{figure}

\clearpage
\captionof{figure}{\textbf{Fixation and exponential growth of two-mutation cells with dedifferentiation for variable stem cell population size.}
A: Observed growth rate $k$ of the stem cell population (black curve) and the semi-analytic approximation Eq.~\eqref{eqn:full k} (green) for $\eta=0$, $\xi=1$, and $u=0.01$. The vertical line denotes $\delta_{crit}$.
B: Analytically predicted critical dedifferentiation rate $\delta_{crit}$ as a function of asymmetric division probability $\eta$ and the growth advantage $s$ of the two-mutation progenitor population.
Exponential growth occurs for $\delta > \delta_{crit}$.
C: Normalized histogram (red crosses) of waiting times for exponential growth of the stem cell population with stochastic homeostasis and dedifferentiation (red) for $u=0.01$, $\delta= 0.01$. For comparison the histogram (red and black dots) as well as the analytical distribution of times to fixation given strict homeostasis for $\varepsilon=0.01$ and $\varepsilon=0$  are also shown.
D: The median and inter-quantile range of times to first occurance of $N_{sc} = 100$ two-mutation stem cells, given stochastic homeostasis and a range of dedifferentiation rates $\delta$. For comparison, the waiting times to fixation for $N_{sc} = 100$ given strict homeostasis (shaded areas) for the equivalent value of $\varepsilon$ are also shown.
E: The probability that the first two-mutation stem cell arose from mutation in the stem cell compartment, rather than dedifferentiation. Vertical line denotes $\delta_{crit}$. Parameters for all simulations given in Table 1. }\label{Figure:dedifferentiation_stats_Moran_variable}

}

	\clearpage
	
	\section*{Tables}
	
	\begin{table}[ht]
	\caption{\bf{Parameter Values Used in Numerical Simulations}}
	\centering
	\begin{tabular}{ |c|c|c|}
	\hline
	\textbf{Parameter} & \textbf{Meaning} & \textbf{Value} \\
	\hline
	$L$ &maximum progenitor lifespan & 20 weeks\cite{Traulsen_bioessays} \\
\hline
	$T_{gen}/N_{sc}$ & mean time between stem cell replication/death events & 0.05 per week \cite{ashkenazi-neoplasia} \\
	\hline
	$N_{sc}$ & total number of active stem cells at homeostasis & 100 \cite{Dingli_Cell_Cycle} \\
	\hline
	$T_{gen}$ & turnover time of active stem cell population & $0.05 \times 100=5$ weeks \\
	\hline	
	$K_{i}$ & active stem cell population carrying capacity & $N_{sc}$ \cite{Dingli_Cell_Cycle} \\
	\hline
	$b_{i}$ & maximal proliferation rate of progenitors with $i$ mutations & 1.5+0.2$i$ per week  \\
	\hline
	$d_{i}$ & maximal death rate of progenitors with $i$ mutations & 1.0 per week  \\
	\hline
	$\rho_b$ & steepness of the proliferation switch & 2 (assumed) \\
	\hline
	$\omega_b$ & age at which proliferation switches off & $10$ weeks\\
	\hline
	$\omega_d$ & age at which death switches on & 0 weeks\\
	\hline
	$u$ & stem cell mutation rate per replication & $10^{-2}$ \cite{Nowak_book} to $10^{-6}$ \cite{araten} \\
	\hline
	$u^*$ & effective progenitor mutation rate (per week) &  $u$ \\
	\hline
$\eta$ & probability of stem cell asymmetric division  & 0 to 1 \\
	\hline
	\multirow{2}{*}{$\varepsilon$} & replicating stem cell fraction from & \multirow{2}{*}{0 to 1} \\
	& dedifferentiated progenitors (constant size model) &  \\
	\hline
	\multirow{2}{*}{$\delta$} & progenitor cell dedifferentiation rate & \multirow{2}{*}{0 to 10 per week} \\
	& (variable size model) & \\
	\hline
	\multirow{2}{*}{$\lambda(t)$} & mean number of dedifferentiated progenitors& \multirow{2}{*}{$\frac{\delta T_{gen} }{S(t)} \int p_2(a,t )\, da$}\\
	& per stem cell reproduction event (variable size model) &\\
	\hline
	\end{tabular}
	\label{parameter_summary}
	\end{table}

\appendix

		\section*{Supplementary Material}
  
  \subsection*{Progenitor cells}  
    \subsubsection*{Analytic solution for progenitor model}
If we neglect mutation, each subpopulation of progenitor cells is governed by a single PDE (Main Text Eq.~1), which can be solved by the method of characteristics.
Without loss of generality, assume
\begin{equation*}
\D{a}{t}=1 \text{ on } \D{p_i}{t}=(\sigma(a)-\mu(a)) p_i.
\end{equation*}
For $a>t_0$ the solution is
\begin{equation}\label{McKendrick_short_time}
p_i(a,t_0)=p(a_0,0) e^{\int_{a_0}^a (\sigma(s)-\mu(s)) ds}.
\end{equation}
The behavior of this solution is determined by the initial condition $p_i(a_0,0)$.
For $a<t_0$ the solution is
\begin{equation}
p_i(a,t_0)=p_i(0,t_0) e^{\int_{0}^{a_0} (\sigma(s)-\mu(s)) ds},
\end{equation}
which results in
\begin{equation}\label{mcKendrick_solution1}
p_i(a,t)=p_i(0,t-a) e^{\int_{0}^{a} (\sigma(s)-\mu(s)) ds}, \quad a<t.
\end{equation} 
Because we are interested in long term behavior, we are only concerned with the solution for $t>a$, which is determined by the boundary condition.


For the stem cell boundary condition (Main Text Eq.~3), the boundary condition does not depend on the state of the population at a previous time, and for $t>a$ equation~\eqref{mcKendrick_solution1} becomes
\begin{equation}\label{mcKendrick_steady_state}
p_i(a,t)=\left(2\alpha_{D,i}+\alpha_{A,i}\right) n_i(t-a) e^{\int_{0}^{a} (\sigma(s)-\mu(s)) ds}.
\end{equation}
Note that the solution has the form $p_i(a,t)=\alpha n_i(t-a) e^{r(a)}$, where $ e^{r(a)}$ is a steady state age distribution, which multiplies the boundary condition provided by the stem cells. Hence, the shape of the age-distribution is largely determined by the functional forms of the the birth and death of the differentiated cells, which are given by Main Text Eq.~5.
(Alternatively Hill functions to model age-dependent proliferation/death rates does not qualitatively change the obtained results and uses the same number of parameters with identical meanings.)

Using equation~\eqref{mcKendrick_steady_state} it is possible to analytically calculate the the steady state age distribution $$e^{\int_{0}^{a} (\sigma(s)-\mu(s)) ds}=e^{r(a)}.$$ For the switch-like birth and death rates given by Main Text Eq.~5 we obtain 
\begin{equation}
r(a)=\frac{(b-d)a}{2}-\frac{b}{2 \rho_b}\left[ \ln\left(\frac{\cosh(\rho_b(a-\omega_b))}{\cosh(\rho_b \omega_b)} \right)\right]-\frac{d}{2 \rho_d}\left[ \ln\left(\frac{\cosh(\rho_d(a-\omega_d))}{\cosh(\rho_d \omega_d)} \right)\right].
\end{equation} 

The progenitor equation (Main Text Eq.~1) can be easily modified to have a maximal carrying capacity $K_i$ for each sub-population.
     \begin{equation}\label{equation:logistic}
	\PD{p_i}{t}+\D{a}{t}\PD{p_i}{a}=\sigma(a)p_i(1-p_i/K_i) -\mu(a)p_i.
	 \end{equation}
     This does not change the qualitative nature of the solutions (See Supplemental Fig.~\ref{FigureS1:Gentry_age_distribution}B,C).

\subsubsection*{Robustness to parameter variations in the progenitor model}\label{section:param_variation}

To test which parameters in the model have the largest effect on the steady state age distribution of differentiated cells, we varied all the parameters in Main Text Eq.~5.
For each parameter the age dependent growth rate $r(a)=\int_{0}^{a} (\sigma(s)-\mu(s)) ds$ and the steady state age distribution $N_{sc}\times  e^{r(a)}$ are plotted in Supplemental Fig.~\ref{FigureS2:growth_rate}.
Note that the proliferation rate $b$ has a great effect on the age distribution of the population.
On the other hand, increasing the removal/clearance rate $d$ does not change the maximal value of the age distribution, only the location at which the peak begins to fall off, because it speeds  up the removal of older cells that are not proliferating.
Increasing $b$ two-fold, from 1 to 2, increases the maximal value of the age distribution 10,000 fold (Supplemental Fig.~\ref{FigureS2:growth_rate}B), but changing $d$  only marginally moves the point at which the age distribution begins to fall off, and does not affect the peak value (Supplemental Fig.~\ref{FigureS2:growth_rate}D).
Note that this result was obtained with the assumption that $\omega_d>\omega_b$, i.e., most cells begin to die off after they are done proliferating.
We also tested the effect of shifting  $\omega_d$ and $\omega_b$.
(Note that $\omega_b$ or $\omega_d$=0 indicates that the rate is constant for all maturity levels.)
A mutation that enables progenitor cells to undergo more divisions before entering senescence and apoptosis has much greater effects on population dynamics than one that enables cell removal to begin earlier.
We can also conclude that enhanced cell clearance rate, whether by the immune system or other methods will not make much difference in this model if the proliferation rate is increased.
Increasing the steepness of either switch ($\rho_b$ and $\rho_d$) did not greatly affect the maximal value of the age distribution, but  made the distribution more box-shaped (Supplemental Fig.~\ref{FigureS2:growth_rate}F,H).

	\subsection*{Alternative Models of Progenitor Dedifferentiation and Competition}

	\subsubsection*{Alternative model for progenitor cells including competition}
The solutions to our progenitor model in the main text are entirely determined by the dynamics of the stem cells, with no interactions between the different mutants.
Here we considered an alternate model including competition between multiple progenitor cell subpopulations.
 Taking into account competition, our PDE system becomes 
 \begin{equation} \label{equation:growth_lotka-Volterra}
\PD{p_i}{t}+\PD{p_i}{a}=(1-u)\sigma_i(a)\left(1-\frac{\sum_{j=0}^M p_i}{K_i}\right)n_i-\mu_i(a)p_i,
 \end{equation} 
where the local carrying capacity of the progenitors in the absence of other subpopulations is $K_i \left(1-\frac{\mu_i}{(1-u)\sigma_i} \right)$.

    We can compare system \eqref{equation:growth_lotka-Volterra} 
      to the classic Lotka-Volterra model. 
  Note that, in general, for a Lotka-Volterra system with multiple species and non-redundant values of reaction constants, it can be shown that  there is only one stable homogeneous equilibrium with one species dominant and the other species extinct (i.e., no co-existence steady state).
However, for PDE system \eqref{equation:growth_lotka-Volterra} we observe coexistence of all three populations for different growth rates. 
  This advection mediated coexistence for competing populations has been previously described for spatial models of competing species \cite{lutscher}, although in our model advection is a maturation process with constant velocity, and no diffusive dispersal takes place. 

Note that in system \eqref{equation:growth_lotka-Volterra}, competition is between individuals of the same maturity stage only. This is appropriate if cells of different maturities can be considered as different cell types or occupy different locations in the body.
If the competition is between cells of all maturity stages, rather than just cells of a given age cohort, then the competition term becomes global rather than  local: \begin{equation}\label{equation:growth_lotka-Volterra_global}
\PD{p_i}{t}+\PD{p_i}{a}=(1-u)\sigma_i(a)\left(1-\frac{\sum_{j=1}^M \int_a p_i(a,t) \, dt }{N_k}\right)n_i-\mu_i(a)p_i(a),
 \end{equation}

 Adding either local or global competition does not influence the proportion of $M$-mutation cells in the progenitor model (Supplemental Fig.~\ref{SupplementalFigure:competition plots}).
We also  looked at the effect of competition between differentiated cells can have on time to cancer acquisition in the full deterministic-stochastic model.
Considering either local or global competition between progenitor subpopulations does not greatly affect time to mutation acquisition in the full model  (Supplemental Fig.~\ref{FigureS3:Moran_waiting_time}B). 
 Hence, without dedifferentiation subpopulation competition in the progenitor model is not significant in altering the time to fixation of mutant in the total cell population for neutral stem cell dynamics.

	\subsubsection*{Alternative model of dedifferentiation for constant $N_{sc}$}
	In the main text, we assumed that only two-mutation progenitor cells could dedifferentiate.
	We also considered the waiting time to fixation when all progenitor cells have  a non-zero probability $\eps$ of dedifferentiating and becoming a cancer stem cell.
	Every $T_{gen}/N_{sc}$ time units, a single randomly chosen stem cell $j$ is removed and one cell $i$ is born with probability given by 
	\begin{equation}\label{equation:dedifferentiation_fixedN_all}
	P(\mathbf{n} \to \mathbf{n}+ \mathbf{e_i} -\mathbf{e_j})=(1-\varepsilon) \frac{n_j}{N_{sc}} \left[
	\sum_{h=0}^{M}m_{h,i} \frac{n_h}{N_{sc}} \right]+ \varepsilon \frac{\int_a p_i(a )\, da}{\sum_i \int_a p_i(a )\, da},\end{equation}
	where $p_i(a)$ is the density of differentiated cells of age $a$ carrying $i$ mutations, and $\varepsilon$ is the proportion of cells in the stem cell pool that come from dedifferentiated cells at each replication event.
		We also considered a model in which all progenitor cells can dedifferentiate, but dedifferentiation is weighted by proliferation rate $\sigma(a)$ of the progenitors, with faster replicating mutants being more likely to end up dedifferentiating. 
	\begin{equation}\label{equation:dedifferentiation_fixedN_all_weighed}
	P(\mathbf{n} \to \mathbf{n}+ \mathbf{e_i} -\mathbf{e_j})=(1-\varepsilon) \frac{n_j}{N_{sc}} \left[
	\sum_{h=0}^{M}m_{h,i} \frac{n_h}{N_{sc}} \right]+ \varepsilon \frac{\int_a \sigma_i(a) p_i(a )\, da}{\sum_i \int_a \sigma_i(a) p_i(a )\, da}.
	\end{equation}
	These assumptions do not significantly change the distribution of waiting times for intermediate dedifferentiation rate $\eps$ (Supplemental Fig.~\ref{FigureS3:Moran_waiting_time}B). However, the waiting time for 2 mutations in this model is faster than model Ib for high values of $\eps$, particularly for small $u$ values (Supplemental Fig.~\ref{FigureS3:Moran_waiting_time}D).


\section*{Derivation of exponential growth rate and critical dedifferentiation rate}
In our variable stem cell population size model, the expected number of two-mutation stem cells produced per stem cell reproduction event is
	\begin{equation}
	\lambda=\frac{\delta T_{gen} }{S(t)} P(t).
	\end{equation}
	A stem cell reproduction event takes place every $T_{gen}/S(t)$ time units, and the stem cell population in changes by
	$
	{\Delta S} =\lambda - (1 - \eta), 
	$
	where $\eta$ is the probability of an asymmetric division.
	Taking the continuum limit, we have
	\begin{equation}\label{eq:S exponential growth}
	\D{S}{t}=\frac{\lambda-(1-\eta)}{T_{gen}} S.
	\end{equation}
	Note that if the mean number of dedifferentiated cells $\lambda$ exceeds $1-\eta$, then $S(t)$ grows exponentially at rate $$k=\frac{\lambda-(1-\eta)}{T_{gen}}.$$
	
	To calculate the growth rate $k$, recall that the two-mutation progenitor population is given by 
	\begin{equation}\label{p_age_distribution}
	p_2(a,t)=\alpha S(t-a)e^{\int_0^a r(x) dx} = \alpha S(t) e^{-k a}e^{\int_0^a r(x) dx}
	\end{equation}
	where $\alpha=2\alpha_{D,2}+\alpha_{A,2}$.
	Here we are making the approximation that most stem cells carry two-mutations, which is valid once exponential growth has proceeded for some time.

	It follows that
	\begin{equation}\label{full_P(t)}
	P(t)=\int_0^\infty p_2(a,t) \, da = \alpha S(t) \int_0^\infty e^{-k a}e^{r(a)}\, da.
	\end{equation}
	Substituting into Eq.~\ref{eq:S exponential growth}, we then have
	\begin{equation}\label{eq:S carrying capacity}
	\D{S}{t}= S(t) \left[ \alpha \delta \int_0^\infty e^{-k a}e^{r(a)}\, da -\frac{(1-\eta) }{T_{gen}}\right].
	\end{equation}

	It follows that $k$ is given by the solution to the integral equation
	\begin{equation}
	k= \alpha \delta \int_0^\infty e^{- k a} e^{r(a)}\, da-\frac{1-\eta}{T_{gen}},
	\end{equation}
	which always has a unique solution for $k$. 

	If
	\begin{equation}
	\int_0^\infty e^{r(a)}\, da > \frac{1-\eta}{\alpha \delta T_{gen}}
	\end{equation}
	then $k=\frac{\lambda-(1-\eta)}{T_{gen}}>0$ and we have exponential growth of the stem cell population. This results in a minimum dedifferentiation rate (per stem cell replication event) necessary for exponential growth given by
	\begin{equation}\label{eq:threshold}
	\delta_{crit} = \frac{1-\eta}{\alpha T_{gen} \int_0^\infty e^{r(a)}\, da}.
	\end{equation}
	This is equivalent to the mean number of cells coming from the dedifferentiated progenitor population being given by
	\begin{equation}\label{eq:threshold_lambda}
	\lambda_{crit} = \frac{(1-\eta)[1+\alpha \int_0^\infty e^{r(a)}\, da ]}{\alpha \int_0^\infty e^{r(a)}\, da}.
	\end{equation}

	This threshold behavior is very similar to what one observes for the well-studied Foerster-McKendrick Equation with an integral boundary condition:
	\begin{align}\label{McKendrick}
	\PD{p_i}{t}+\PD{p_i}{a}&=-\mu(a)p_i,\\
	p(t,0)&=\int_0^\infty b(a)\, p(t,a)\, da.
\end{align}
	There, the behavior of the solution depends on a critical quantity
	\begin{equation*}
	s=\int_0^\infty b(a) \exp \left(-\int_0^a \mu(s)\, ds\right)\, da.
	\end{equation*} The solution undergoes exponential growth if $s>1$, and exponential decay if $s<1$ \cite{Murray_book}.
	Our model, however, has a mixed boundary condition for $p_2(t,a)$.
	Hence, when $\delta< \delta_{crit}$ and $k<0$, the stochastic term in the boundary condition dominates, and we get similar behaviour as the variable $S(t)$ case with no dedifferentiation. That is, there is neither exponential growth nor decay, and the waiting time distribution to fixation is equivalent to the fixed N case with $\delta=0$. 
	
	\section*{Supplemental Figures}
	
	    \begin{figure}[htbp]
	\begin{centering}
	\includegraphics[width=17.35cm]{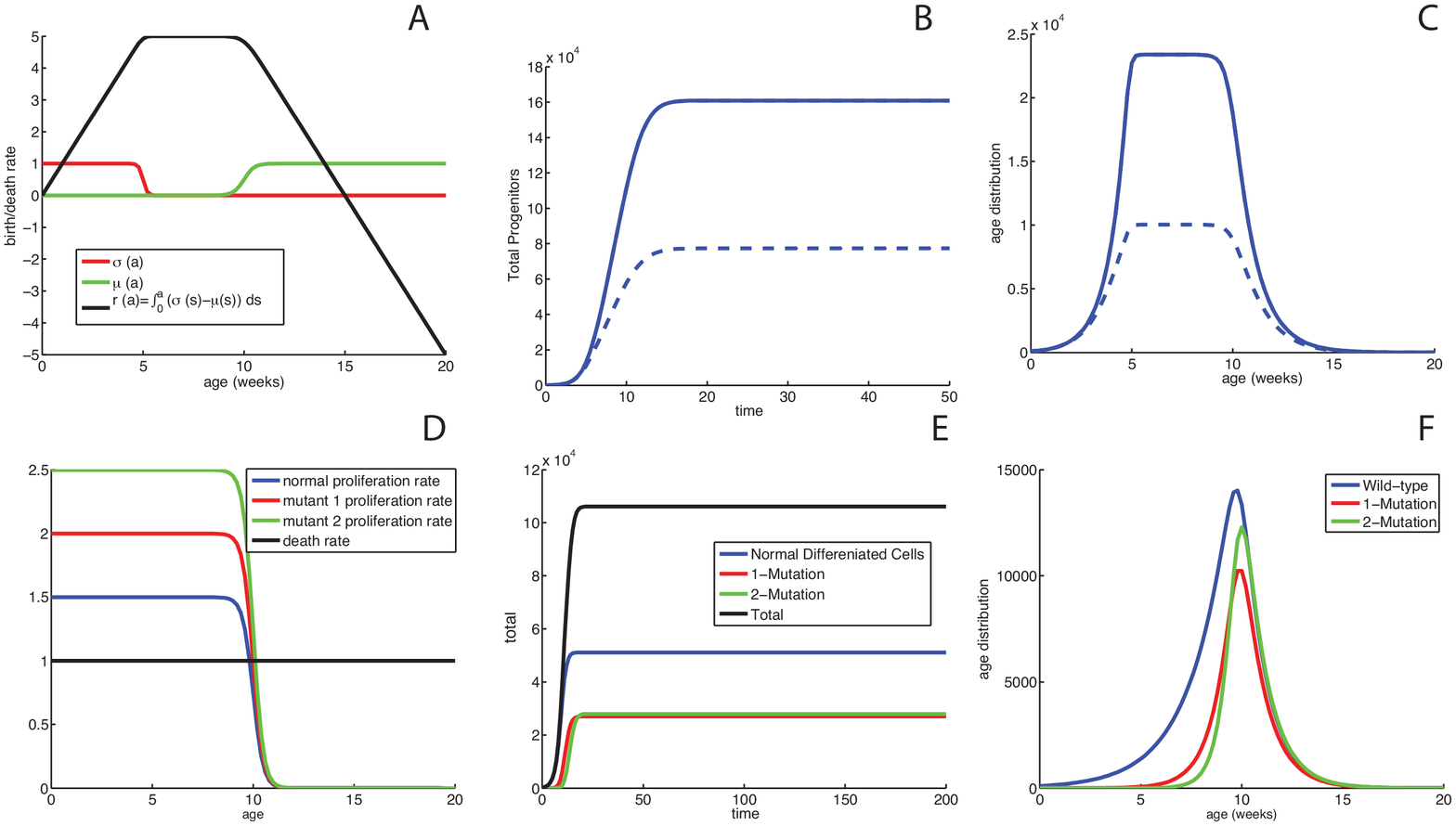}
	\par\end{centering}
	\caption{ \textbf{Progenitor cell dynamics for a constant stem cell population.} 
    Fixed number of zero-mutation stem cells $N_{sc}=100$. (A) Proliferation and death rates $\sigma(a)$ (red curve) and $\mu(a)$ (blue curve) used in the simulation and the resulting age-dependent growth rate $r(a)$ (black curve). (B) The total number of  progenitor cells  given by (Main Text Eq.~1) (solid line) and equation \eqref{equation:logistic} (dotted line)  as a function of time. 
 (C) The steady state age distribution with no competition between progenitor cells is given by (Main Text Eq.~1)  (solid line) and with logistic growth by equation \eqref{equation:logistic} (dotted line).
	(D-F) Progenitor dynamics for for $K=3$ subpopulations with mutation rate $u=0.001$.
	(D) Age-dependent birth rates $\sigma(a)$ 
	given by Main Text Eq.~5. The death rate is constant. (E) The total number of progenitor cells as a function of time for zero-mutation (blue), one-mutation (red), and two-mutation (green) subpopulations.  (F) The steady state age distribution for zero-mutation (blue), one-mutation (red), and two-mutation (green) progenitor cells. 
	}
	\label{FigureS1:Gentry_age_distribution}
	\end{figure}
	
		\begin{figure}[htbp]
	\begin{centering}
	\includegraphics[width=17.35cm]{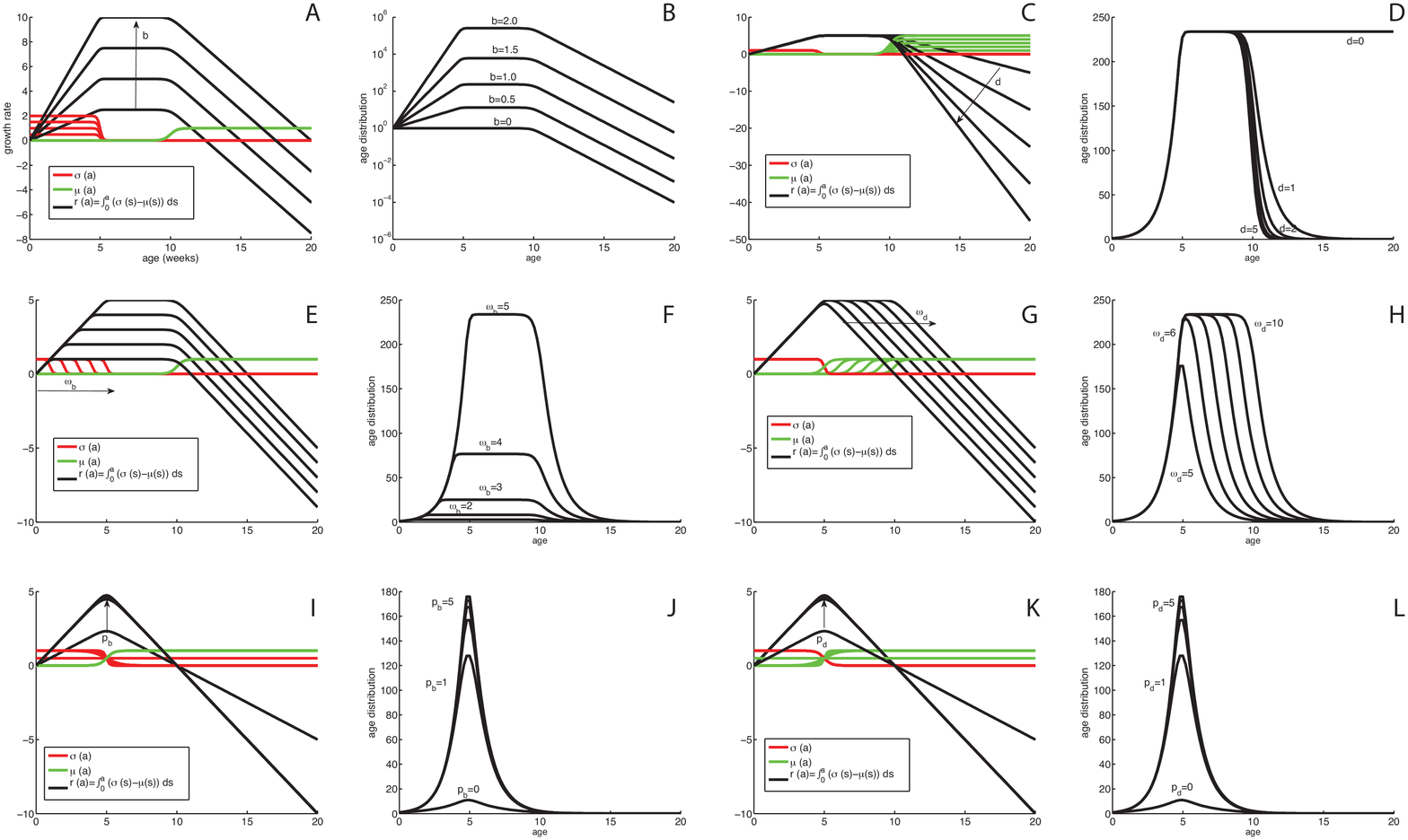}
	\par\end{centering}
	\caption{\textbf{ Robustness to parameter variations in proliferation/death rates in progenitor model.} 
	The effect of parameters in age-dependent birth and death rates, $\sigma(a)$ and $\mu(a)$ (given by Main Text Eq.~5) on the age-dependent growth rate $r(a)=\int_{0}^{a} (\sigma(s)-\mu(s))\, ds$ and the steady state age distribution $e^{r(a)}$ in our models.
	(A,B) Effect of varying maximal growth rate $b$ between 0 and 2. (C,D). Effect of varying maximal death rate $d$ between 0 and 5. (E,F) Effect of varying the location (age of onset) of the proliferation switch $\omega_b$ between 0 and 5. (g,h) Effect of varying the age at which the apoptosis switch $\omega_d$ is turned on between 5 and 10. (I,J) Effect of varying the steepness of the proliferation switch $\rho_b$ between 0 and 5. (K,L) Effect of varying the steepness of the apoptosis switch $\rho_d$ between 0 and 5.}     \label{FigureS2:growth_rate}
	\end{figure}
    
    \begin{figure}[htbp]
\begin{centering}
	\includegraphics[width=17.35cm]{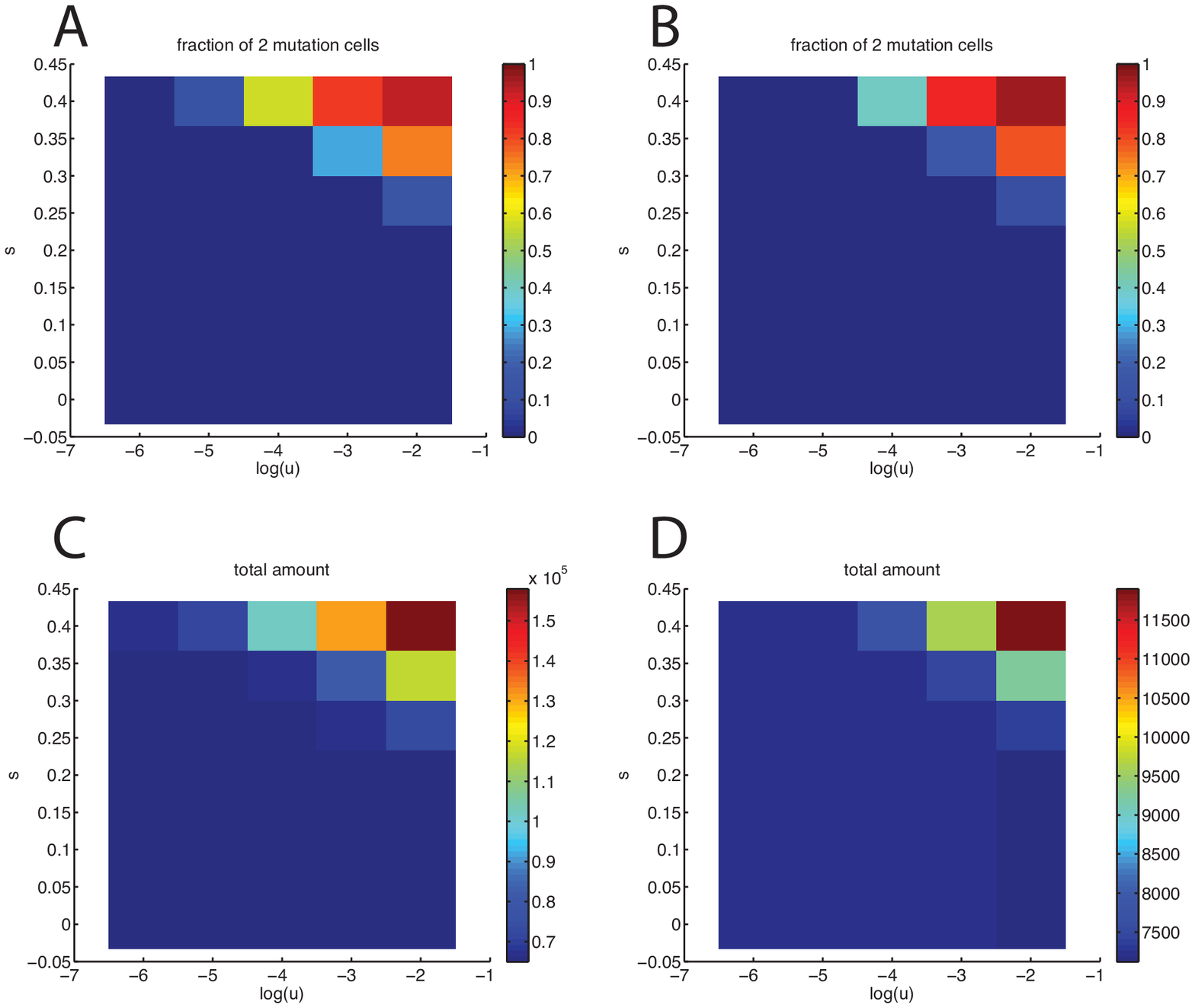}
\par\end{centering}
\caption{ \textbf{Steady-state progenitor distributions in the absence of stem cell mutation  but with progenitor competition.} Top: The fraction of mutant cells as a function of mutation rate $u$ and proliferative advantage $s$ for (A,C)  local (age-dependent) competition between subpopulations given by equation~\eqref{equation:growth_lotka-Volterra}, and (B,D) global competition between subpopulations given by equation \eqref{equation:growth_lotka-Volterra_global}.  Bottom: Corresponding plots of total cell density. Basal dynamics are constant death rate $\mu=1$ and sigmoidal birth rate with maximal growth rate $b_0=2$, $b_i=(1+s)b_{i-1}$ for $i=1,2$. The same carrying capacity is used for all simulations: $N_1=200 N_{sc}$, $N_2=250 N_{sc}$, $N_3=300 N_{sc}$. Note that there is a sharp transition zone at which mutant cells go from nearly zero fraction of total population to majority of the differentiating cell population.  However, the mutation rate $u$ and proliferative advantage $s$ at which this is observed is unreasonably high, just as for the model without progenitor competition (Main Text, Fig ~2).}
\label{SupplementalFigure:competition plots}
\end{figure}
	
	\begin{figure}[htbp]
	\begin{centering}
\includegraphics[width=17.35cm]{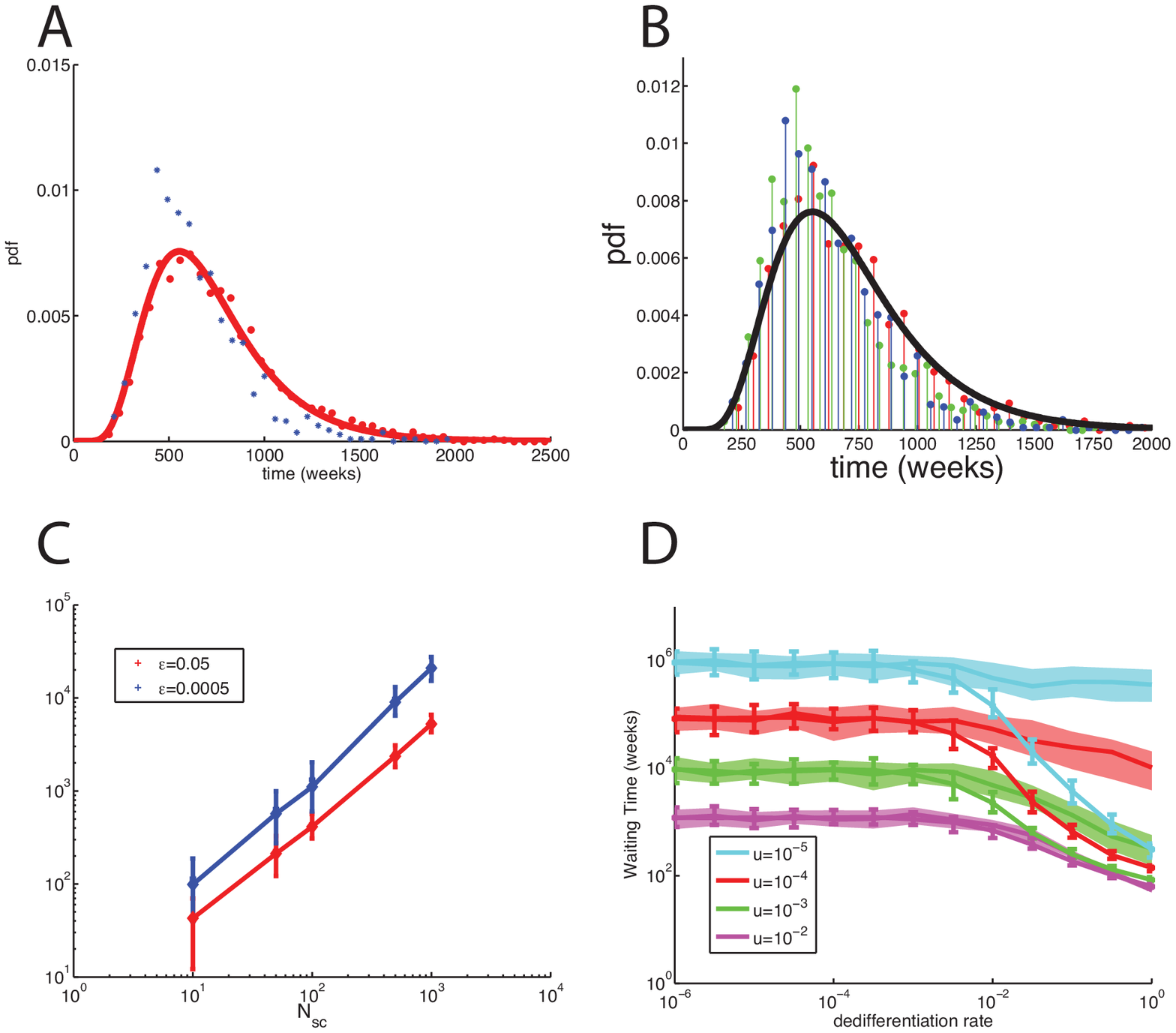}
	\par\end{centering}
	\caption{\textbf{Comparison of two Model I variants with all-mutant  progenitor dedifferentiation and two-mutant progenitor dedifferentiation.} 
	(A) Fixation time distributions in constant stem cell population size model for potential dedifferentiation of only two-mutation progenitors (red, Main Text Eq.~8) and potential dedifferentiation of all progenitor cells (blue, Eq.~\ref{equation:dedifferentiation_fixedN_all}).  
	(B) Fixation time distributions in constant stem cell population size model with  dedifferentiation of all progenitor cells.  
	 Blue: all progenitor cells equally likely to dedifferentiate with dedifferentiation probabilities given by \eqref{equation:dedifferentiation_fixedN_all}).   Red:  all progenitor cells can dedifferentiate with dedifferentiation probability weighed by birth rate given by~\eqref{equation:dedifferentiation_fixedN_all_weighed}.  Progenitor dynamics without competition (Main Text Eq.~2).  Green: all progenitor cells can dedifferentiate with dedifferentiation probability weighed by birth rate given by~\eqref{equation:dedifferentiation_fixedN_all_weighed}. Progenitor dynamics with local competition given by~\eqref{equation:growth_lotka-Volterra}.
	 Dedifferentiation rate used is $\varepsilon=0.02$, mutation rate is $u=0.01$. 
	 (C) Mean $\pm$ standard deviation of time to fixation as the stem cell pool size $N_{sc}$ is varied for two different values of the dedifferentiation rate $\eps$. Mutation rate is $u=0.01$. 
	 (D) Median and inter-quantile range of time to fixation in alternative Model Ib as a function of dedifferentiation rate $\epsilon$ are shown as a box-whiskers plot. All mutant cells are allowed to dedifferentiate with probability of dedifferentiation give by eq.~(\ref{equation:dedifferentiation_fixedN_all})  $u=0.01$ (blue), $u=0.001$ (green),$u=0.0001$ (red), and $u=10^{-5}$ (teal). For comparison, the waiting times to fixation in Model Ib  are also shown as shaded areas (compare to Main Text, Fig. 4C).
}\label{FigureS3:Moran_waiting_time}
	\end{figure}

		\begin{figure}[htbp]
	\begin{centering}
\includegraphics[width=17.35cm]{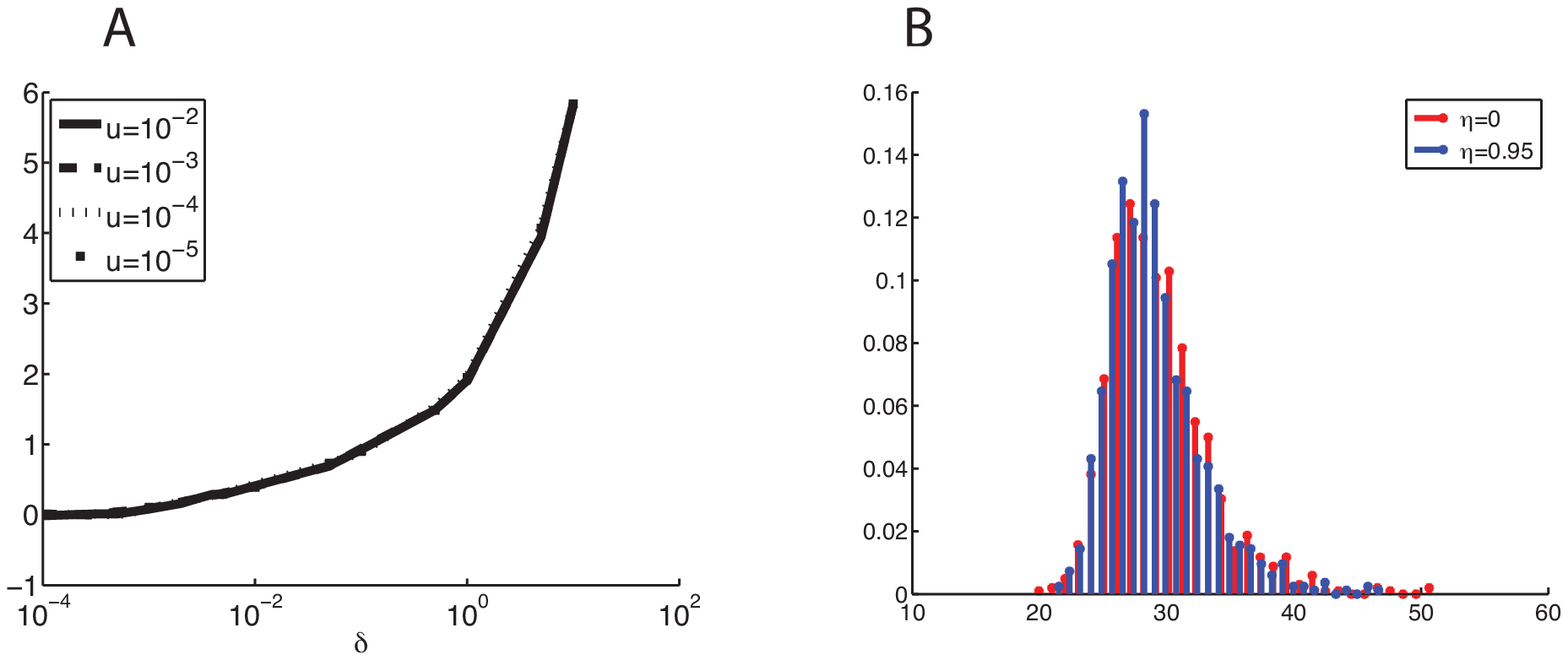}
	\par\end{centering}
	\caption{ \textbf{Supplemental Fig. 5. Characterization of exponential growth of two-mutant population in Model II.} (A) 	The exponential growth rate $k=\frac{\lambda-(1-\eta)}{T_{gen}}$  of the stem cell population does not depend on the mutation rate ($u=0.01, 0.001, \cdots, 10^{-5}$ for $\eta=0, \, \xi=1$).
	(B) The time to exponential growth for different rates of asymmetric division (red $\eta=0$; blue:$\eta=0.95$) is roughly similar. Rate of dedifferentiation is $\delta=0.01$.  $n=1000$ points are used for each distribution.
}\label{FigureS4:Variable N}
	\end{figure}

	\end{document}